\documentclass[12pt]{article}

\usepackage{epsfig}

\def\be{\begin{equation}}
\def\ee{\end{equation}}
\def\ba{\begin{eqnarray}}
\def\ea{\end{eqnarray}}

\begin{document}

\thispagestyle{empty}
\def\thefootnote{\fnsymbol{footnote}}
\begin{flushright}
 DAMTP-2000-121 \\
 RUNHETC-2000-41 \\
 hep-th/0010247 \\
 \end{flushright}
\vskip 0.5cm

\begin{center}\LARGE
{\bf Renormalization Group Analysis}\\
{\bf of Tachyon Condensation}
\end{center}

\vskip 1.0cm

\begin{center}
{\centerline{Satabhisa Dasgupta\footnote{{\tt 
satavisa@physics.rutgers.edu}}}}

\vskip 0.5cm

{\it Department of Physics and Astronomy, Rutgers University,
\\ Piscataway, NJ 08855, U.S.A.}
\end{center}

\vskip 0.5cm

\begin{center}
{\centerline{Tathagata Dasgupta\footnote{{\tt 
T.Dasgupta@damtp.cam.ac.uk}}}}

\vskip 0.5cm

{\it Department of Applied Mathematics and Theoretical Physics
\\ Wilberforce Road, Cambridge, CB3 0WA, U.K.}
\end{center}

\vskip 1.0cm


\begin{abstract}

Renormalization group analysis of boundary conformal field theory on
bosonic D25-brane is used to study tachyon condensation. Placing the
lump on a finite circle and triggering only the 
first three tachyon modes, the theory flows to nearby IR fixed point
representing     
lumps that are extended object with definite profile. The boundary
entropy corresponding to the 
D24-brane tension is calculated in the leading order in perturbative
analysis which decreases under RG flow and agrees with the expected result
to an accuracy of $8\%$. Multicritical   
behaviour of the IR theory suggests that the end point of the flow
represents a configuration of two D24-branes. Analogy with Kondo
physics is discussed.  We comment on
$U(\infty)$ symmetry restoration.

\end{abstract}

\vfill
\setcounter{footnote}{0}
\def\thefootnote{\arabic{footnote}}
\newpage


\section{Introduction}
\setcounter{equation}{0}

It has been suggested by Sen~\cite{sen-univ} that the energy gap
between unstable D-brane configurations and the stable vacuum should
be computable using Witten's cubic string field
theory (SFT)~\cite{witten-86}. The level truncation scheme
of~\cite{kost-sam} has appeared to lead to very good agreement with the
expected results in the context of the decay of bosonic
D-branes~\cite{sen-zwiebach,moeller-taylor} and unstable D-branes in
superstring field
theory~\cite{berkovits,berk-sen-zwie,smet-rae,iqbal-naqvi}.    
Lower dimensional D-branes have been constructed as tachyonic lump
configurations on bosonic
D25-brane~\cite{kjmt,harvey-kraus,msz,koch-rodrigues,moeller}. 
Support for Sen's conjecture that lower dimensional D-branes can be
identified with tachyonic lump solution of string field theory on
bosonic D25-brane also has come from the noncommutative limit of the
effective field
theory~\cite{gopa-min-stro,das-mukhi-raj,har-kraus-lar-mar,witten-tachy}.
Various toy models of tachyon dynamics have been useful tool for 
understanding the realization of Sen's
conjectures~\cite{zwiebach-toy,min-zwie/1,min-zwie/2}.

Despite the success of cubic SFT in leading to very good agreement with the
expected results, it is still not clear why the calculations in the
level-truncated string field theory converge so rapidly to the correct
answer for quantities like vacuum energy. It is also not clear how to
study the nonperturbative vacuum using this approach. An alternative
method has been suggested in~\cite{hkm} which says that
renormalization group (RG) analysis of worldsheet theory in first
quantized approach can be used to show that the mass of the tachyon
lump on a Dp-brane is equal to that of a D(p--1)-brane. This leads to
the idea that boundary string field theory (SFT) as was originally 
proposed by Witten and
Shatashvili~\cite{witten-bsft1,witten-bsft2,shat2,shat1} may
efficiently describe 
open string tachyon condensation on D-branes 
in bosonic string theory~\cite{ger-shat,kmm,ghoshal-sen/norm} (also
see~\cite{and}. 
It has been pointed out in~\cite{ger-shat,kmm} that boundary SFT can
provide an exact verification of Sen's conjectures. Based
on~\cite{witten-bsft2,shat1}, one can compute the action exactly
taking a simplest tachyon profile as boundary operator of ghost number
zero which is quadratic in the space-time coordinate. One can also
describe the lumps corresponding to the lower dimensional D-branes and
calculate their tension that agrees with the expected result exactly. 

One strong point of the results of~\cite{msz} using level truncation
scheme in cubic SFT is that it gives a definite picture of the
tachyon profile as superposition of $\cos (\frac{n}{R}X)$ for
different $n$ with definite coefficients producing soliton of finite
width. On the other hand the boundary SFT analysis of~\cite{kmm} does not say 
how the higher tachyon harmonics get mixed with the cosines of
different $n$ to produce finite size soliton profile. The reason for this
is that the mass parameter flows to infinity in the IR in their
particular choice of coordinates starting with the simple choice of
the tachyon profile initially. This is because the RG equations in
that particular choice of coordinates become linear in coupling
constants. Since the width of the soliton is given
by inverse of the mass parameter which flows to infinity in the exact
description, the width is zero. Also it is difficult to see in this
setup how the theory at IR fixed point can describe configurations of
more than one D-brane. We should note that although the theory flows
to infinity in the IR, the Zamolodchikov metric on the space of
worldsheet theories measures the distance between the UV and IR fixed
point to be finite. 

In this paper we will try to address these issues in the worldsheet
approach by choosing the initial tachyon profile on a circle on 
bosonic D25-brane world-volume. The basic setup is similar
to~\cite{msz}. The choice of boundary perturbation, as we will 
discuss in the next section, is motivated from the analysis
of~\cite{msz}. In our choice of coordinates in the space of worldsheet
field theories the RG analysis provides the boundary conformal field theory to
flow to a nearby IR fixed point. We will consider only the first
three tachyon modes. The values they flow to after RG flow appears to
be in good agreement with the values computed in~\cite{msz} of tachyon
modes of the string field at the stationary point of the potential.

The plan of the paper is the following. In section 2 we briefly review
the worldsheet RG scheme in the context of boundary SFT. We
comment on the boundary entropy that measures the corresponding
D-brane tension describing the boundary CFT. We highlight
the basic setup of~\cite{msz} that leads to a particular choice of the
tachyon profile. 

In section 3 we perform the RG analysis in detail choosing a
particular ansatz for the Green's function of fast moving modes of the
scalar field. The RG equations are obtained up to third order in
coupling constants. The easy but tedious parts of the calculations are 
given in the appendix. 

In section 4 we plot the lump profile that appears from the RG
analysis of the previous section. Following the method of Affleck and
Ludwig~\cite{al-prb} we calculate the boundary entropy in the leading
order. Our result satisfies the $g$-theorem of~\cite{al-prb}. The
boundary entropy is calculated to an accuracy of about $8\%$ of the
exact result.  

In section 5 we analyse the multicritical behaviour of the potential
in the IR and argue that the theory in the IR is that of a
configuration of two D24-branes. We make an analogy with the Kondo
problem and argue that the process of the formation of the lump due to
tachyon condensation corresponds to the underscreened Kondo effect. 
Exact screening occurs when the theory on the brane rolls down to the
nonperturbative closed string vacuum. Overscreening, on the other
hand leads to a picture similar to dielectric effect. We comment on
$U(\infty)$ symmetry restoration. 

Section 6 contains discussions on further related issues that are
beyond the scope of the paper.

\newpage


\section{Open string in tachyon background}
\setcounter{equation}{0}

If the spectrum of a point in the moduli space has relevant operators in the
IR, the point is said to be unstable. The unstable point then might be 
discarded from the moduli space. 
In other words, this results in the appearance of unstable 
directions in the effective potential. IR instability indicates that we are
in a {\it false} vacuum. The obvious question will be 
which point (or vacuum) will replace this unstable point, or in other words,
how to resolve the IR instability. 

Often regions in moduli space contain 
D-brane configurations related by T-duality. The above issue can be addressed
in these sectors by adding some IR relevant boundary perturbations.
One important feature of this boundary deformation is that the bulk theory
always remain conformal. Flows caused by a relevant boundary operator 
appear as open string tachyon condensation. Flowing to IR on the worldsheet
is equivalent to approaching a classical solution of spacetime theory. 

As a result of the flow, at some points in moduli space, boundary conditions
are changed from Neumann to Dirichlet. The reverse process indicates
nonunitarity. The direction of the flow is determined by 
{\it boundary entropy} defined by~\cite{al-prl} 

\be 
g_a = \langle 0|a\rangle\,,
\ee
the disk partition function of the boundary state $|a\rangle$ associated
to the perturbed theory. The phases of $|0\rangle$ and $|a\rangle$ can
be chosen such that the above quantity is real and positive for any 
boundary state. It is shown in~\cite{al-prb} 
in first order in conformal perturbation theory that boundary entropy, 
$g$ decreases along RG flows, suggesting the so called $g$-{\it theorem}
similar to Zamolodchikov's $c$-theorem~\cite{zamo-c}. In this context,
the case of Sine-Gordon boundary perturbation with single frequency 
is studied in~\cite{fsw-exact,fls-plasma}. 

In fact $g$ measures the tension of the D-brane that describes boundary 
conformal field theory at the corresponding fixed point~\cite{hkms} 
(see also~\cite{el-rab-sar}) and the 
$g$-theorem implies the minimization of the action in the space of 
{\it all two-dimensioanal worldsheet field theories}. The worldsheet
formulation 
of string field theory is {\it manifestly} background
independent\footnote{Of course, here we are referring to the open
string background. The definition of $Q$ depends on the closed string
background.}. In~\cite{witten-bsft1} a gauge invariant background   
independent spacetime string field theory action is defined as the
solution of the following equation:

\be
\frac{\partial {\cal S}_{\mbox{bsft}}}{\partial \lambda_i} = \frac{K}{2}
\int_0^{2\pi}\frac{d\theta}{2\pi}\int_0^{2\pi}\frac{d\theta'}{2\pi}
~\langle {\cal O}_i(\theta )\{Q,{\cal O}(\theta')\}\rangle_\lambda\,,
\label{Seqn}
\ee
where $Q$ is the BRST charge and and $\theta$ is the boundary parameter of
the disk. The operator ${\cal O}$ has ghost number one: ${\cal O} = c{\cal
V}$, where ${\cal V}$ is a general boundary perturbation describing the
space of boundary perturbations: ${\cal V} = \sum_i \lambda^i{\cal
V}_i$. The normalization constant $K$ is fixed  
in~\cite{ghoshal-sen/norm} by comparing the on-shell three tachyon amplitude
computed from the background independent open string field theory with
the same computed in Chern-Simons string field theory:

\be
K = T_p\,,
\label{norm}
\ee
$T_p$ being the corresponding Dp-brane tension. The correlation function
in the above expression is defined in terms of the perturbed worldsheet 
action on the disk. The above action is defined up to an additive
constant. Although the boundary perturbation ${\cal V}$ does not depend
on ghosts, the theory with perturbed action is not renormalizable. But
in the case of tachyon condensation, it is renormalizable. So boundary
string field theory works well for studying tachyon condensation
problem. But it is not clear how to define a general off-shell
amplitude due to UV divergences on the worldsheet. 

An alternative definition of {\cal S} is given by  
the metric on the space of worldsheet theories~\cite{kmm}

\be
\frac{\partial {\cal S}_{\mbox{bsft}}}{\partial \lambda_i} = -\beta^jG_{ij} \,,
\label{S-G}
\ee
where $\beta_i$ is the corresponding beta function which acts like a 
vector field on the space of worldsheet theories. In~\cite{witten-bsft2,
shat1} a very important relation between the action ${\cal S}$ and the 
partition function on the disk is demonstrated up to second order in
coupling constant:

\be
{\cal S}_{\mbox{bsft}} = -\beta^i\frac{\partial Z(\lambda )}
{\partial \lambda^i}+Z(\lambda )\,.
\label{s-z}
\ee
Since on-shell the beta function vanishes, the action is same as the partition
function. This fixes the additive ambiguity in~(\ref{Seqn}). 
The above relation implies that all symmetries of the worldsheet partition
partition function are also symmetries of the spacetime action.
It is further argued that~\cite{kmm} the spacetime 
action ${\cal S}$ is nothing but the boundary entropy $g$. Although
background independence is manifest 
in the worldsheet formalism, it is lost once we compute ${\cal S}$ or
$g$ perturbatively. But if the above relation between the spacetime action
and the partition function is true in all orders in coupling constant, it is
not lost. 

On the other hand, the Chern-Simons string field theory is not manifestly
background independent. To achieve this one is forced to work in a truncated
version of Hilbert space of the first quantized theory restricting the string
field to a background independent subspace for studying the classical 
lump solution. In our renormalization group analysis of the problem of 
formation of the tachyonic lump, we follow the basic setup
of~\cite{msz} where the $x_{25}$ coordinate is taken to be on a circle of 
radius $R_{25}$. We take $\Phi$ to be the scalar field on the string 
worldsheet associated with $x_{25}$. On a D25-brane $\Phi$ 
satisfies Neumann boundary conditions. The conformal field theory
associated with the field $\Phi$ has central charge, $c=1$. 

Before going into the RG analysis let us briefly discuss the setup
of~\cite{msz}. For details on the background independent subspace of the
complete Fock space see~\cite{sen-univ}. A basis of states
of this theory is conveniently formed by grouping the states into Verma
modules. The states contained in the Verma module are obtained out of primary
state $\exp (in\Phi (0)/R_{25})|0\rangle$ by acting with the associated
Virasoro generators $L^{\Phi}_{-m}$. Linear independence of the states to
form the basis is achieved by removing null states from the spectrum.
In fact the spectrum is free of null states for $n \neq 0$ and an irrational 
$R_{25}$ value. In order to achieve a successful truncation of the Hilbert
space one has to restrict the primary states of the boundary conformal field 
theory along $x_{25}$ to be even under $\Phi \rightarrow -\Phi$ and to be
trivial primaries of the conformal field theory of the fields associated to
the rest of the coordinates of bosonic string theory. As a result the 
appropriate primary states are: (1) the zero momentum primaries that are
even under $\Phi \rightarrow -\Phi$ (also removing the null states), and 
(2) the vacuum state $\cos (n\Phi (0)/R_{25})|0\rangle$ with $n\neq 0$. 

Let us recall that an open bosonic string propagating in the tachyon background
is described by the following action

\be
S=\frac{1}{4\pi}\int dsdt~\eta^{ab}\partial_a\Phi_\mu\partial_b\Phi^\mu
+\int_{-\infty}^{\infty} \frac{dt}{a}\int dk~T(k) e^{ik\Phi}\,,
\ee
where $a$ is the UV cutoff and $T(k)$ is the tachyon field with momentum
$k$. In the present case where $x_{25}$ is compact, the momentum is
descrete and we are motivated to take the boundary perturbation to be
of the form (respecting $\Phi \rightarrow -\Phi$ symmetry):

\be
\int dt~T[\Phi (t)]=\sum_{n=0}^\infty\lambda_n\int dt~\cos\Big(\frac{n}{R_{25}}
\Phi (0,t)\Big)\,, 
\ee
where $\lambda_n$s are tachyon modes. The zero mode $\lambda_0$ is the
identity operator and is always tachyonic. The higher modes may not be
tachyonic depending on the radius $R_{25}$. In our analysis we only consider
the first three tachyon modes ({\it i.e.} $n=0,1,2$). Hence the problem is
similar to the boundary Sine-Gordon model with two frequencies, which
typically allows to hit a nearby IR fixed point producing a lump
profile of finite size.


\section{Switching on first three tachyon modes: RG analysis}
\setcounter{equation}{0}

The boundary conformal field theory can be described by a Gaussian 
model with gapless excitation spectrum and the correlation functions of 
bosonic exponents follow power laws. This behaviour implies that the 
correlation length is infinite and the system is in its critical 
phase. Under boundary perturbations, usually the correlation functions 
are affected differently on different scales and the long distance 
asymptotics get affected the most. Certain perturbations may cause only tiny 
changes in the UV, but changes the IR behaviour profoundly. In the
RG picture this is observed as a growth of the coupling constant
associated with the perturbation. A slow decay of correlation functions 
gives rise to divergences in the perturbation series.

If the influence of the perturbing operator grows on large scales, the
perturbation is {\it relevant}. Consider the perturbed action 

\be
S = S_0 + \lambda\int dt {\cal O}_\Delta (t)
\ee
where $S_0$ is the action of the system at criticality and 
${\cal O}_\Delta (t)$ is the perturbing field with scaling dimension 
$\Delta$, $\langle {\cal O}_\Delta (t_1){\cal O}_\Delta^+ (t_2)
\rangle \sim |t_1-t_2|^{-2\Delta}$. The perturbation with zero 
conformal spin and scaling dimension $\Delta$ is relevant if $\Delta<1$
and {\it irrelevant} if $\Delta>1$. The case $\Delta = 1$ is the 
{\it marginal} one and its effect on scaling dimensions of correlation
functions depends on the sign of the coupling constant $\lambda$.

Consider the path integral on the disk with a worldsheet action 

\be
S = S_0 + \int dt~T[\Phi(t)]
\label{totalaction}
\ee
where $S_0$ is the free field action on the disk describing an open and closed
string conformal background. The perturbing boundary field $T[\Phi(t)]$ is a 
tachyon profile and $\Phi$ is the scalar field along the string worldsheet 
associated with the compactified coordinate $x_{25}$. The above action 
describes a renormalization group flow from a theory where all 26 bosonic
directions are Neumann describing a D25-brane to the theory where $x_{25}$
direction is Dirichlet describing a D24-brane. 

Instead of using the Callan-Symanzik formalism 
we follow the Wilsonian renormalization scheme where we put the theory 
on a lattice. 
We assume that the Fourier transform of the field, 
$\Phi(\omega)$, is defined in the Brillouin zone and choose a cut-off,
$\omega < \Lambda$. The aim is to start moving towards larger distances
(or lower energies) by integrating out the fields with shorter and shorter
wavelengths. The procedure is based on the decomposition of the boundary 
field with a cutoff into
a slow moving (long-wavelength) and a fast moving (short-moving)
component, $\Phi_\Lambda=\Phi_{s\Lambda'}+\Phi_f$, where we have split the 
Brillouin zone $\omega < \Lambda$ into a wide region 
$0<\omega<\Lambda'=\Lambda-d\Lambda$ and a narrow slice
$\Lambda'=\Lambda-d\Lambda<\omega<\Lambda$. The original field is given by

\ba
\Phi_\Lambda (t) &=& \int {d\omega \over 2\pi} e^{i\omega t}
\Phi_\Lambda(\omega) \nonumber \\ 
&=& \int_{|\omega|<\Lambda'} {d\omega \over 2\pi}
e^{i\omega t} \Phi_{s\Lambda'}(\omega) +
\int_{\Lambda'<|\omega|<\Lambda} {d\omega \over 2\pi}
e^{i\omega t} \Phi_f(\omega) \, .
\ea
The next step will be to perform a partial path integration in the partition 
function over the fast moving part and representing the result in terms of 
an effective model for the slow moving field. If the model is renormalizable,
which is known to be the case for Sine-Gordon model that we will consider
below, the effective action will have the same structure as the original one, 
but with a new set of coupling constants. This procedure is repeated several 
times, and after each step the form of the original model is reproduced
(upto irrelevant terms). The relations between the bare and the renormalized
couplings then lead to renormalization group (RG) equations.

The Gaussian part $S_0$ is additive under the above decomposition. Hence, 
given the cutoff, the partition function is given by 

\ba 
Z_\Lambda &=& \int
D\Phi_{s\Lambda'}D\Phi_f~e^{-S_0[\Phi_{s\Lambda'}]}
e^{-S_0[\Phi_f]}e^{-S_I[\Phi_{s\Lambda'}+\Phi_f]} 
\nonumber \\ 
&=&Z_f\int D\Phi_{s\Lambda'}~e^{-S_0[\Phi_{s\Lambda'}]}
\langle e^{-S_I[\Phi_{s\Lambda'}+\Phi_f]}\rangle_f 
\ea
where $Z_f$ is a nonsingular contribution of fast moving components 
to the partition function. The effective action involving the slow moving
part of the field is given by

\be
S_{eff}[\Phi_{s\Lambda'}]=S_0[\Phi_{s\Lambda'}]+S_{I~eff}[\Phi_{s\Lambda'}]
\ee
where

\be
S_{I~eff}[\Phi_{s\Lambda'}] 
= -\ln\langle e^{-S_I[\Phi_{s\Lambda'}+\Phi_f]}\rangle_f 
\ee
It is clear that the effective action preserves all IR singularities.
Assuming the coupling to be small, we expand the above effective interaction
perturbatively upto cubic order 

\ba
S_{I~eff}[\Phi_{s\Lambda'}]
&=& \langle S_I \rangle_f+{1 \over 2}(\langle S_I\rangle_f^2-\langle S_I^2 
\rangle_f) 
\nonumber \\
& &+({1 \over
6}\langle S_I^3\rangle-{1 \over 2}\langle S_I\rangle_f \langle S_I^2 \rangle_f
+{1 \over 3}\langle S_I\rangle_f^3) 
+O(\lambda^4)
\label{SIeff}
\ea

Now we consider the effect of switching on first three tachyon
harmonics: $\lambda_0,~\lambda_1$, and $\lambda_2$. The scalar
perturbation is of {\it double Sine-Gordon} type and reads
\footnote{One can take $(\pm 1)^n$ symmetry of the coupling $\lambda_n$s
in writing down the perturbation which might be more appropriate in 
order to compare the results 
with that of~\cite{msz}. However we chose perturbation of the type~(\ref
{double-sine-gordon}) as this allows us to hit the desired multicritical
IR fixed point from the stable direction. However we will
invoke the $(-1)^n$ symmetry to show multicriticality in section 5.
The opposite sign in front of the identity operator does not modify the
analysis for other operators. The coupling in front of the identity 
operator appears linearly in the beta function and it does not mix with
other couplings. Also as we will see in the next section, boundary 
entropy does not get a contribution from it.} 

\be
S_I[\Phi_\Lambda] = \lambda_0\int dt~{\bf 1} 
- \lambda_1\int dt~\cos\beta\Phi_\Lambda(t) 
- \lambda_2\int dt~\cos 2\beta\Phi_\Lambda(t) \,.
\label{double-sine-gordon}
\ee  
where $\beta$ defines the conformal dimensions of our theory and
the scaling dimension is $\Delta_1 = \beta^2/4\pi$ for the first harmonic 
and $\Delta_2 = \beta^2/\pi$ for the second harmonic. The tachyon zero 
mode is the identity operator with zero scaling dimension. 
Rescaling $\beta$: $\beta \rightarrow \sqrt{4\pi}\beta$ results in the relation
$\Delta_n=n/R_{25}$. Following~\cite{msz}, we take the radius
$R_{25}=\sqrt 3$.
In this radius the first mode is least relevant and the perturbation
by the second 
mode is least irrelevant, which, after being added
to the cutoff theory, should improve the RG result. Such a procedure
should lead to a good result even when only first few irrelevant terms
are included. The effects of highly irrelevant operators are highly
damped by their rapid decay. 

Using the above definition of the effective action we perturbatively
calculate contributions in each order in coupling constant,
$\lambda$. Then in order to restore the original cut-off to $\Lambda$
we rescale the 
energy $\omega'=(\Lambda/\Lambda')\omega \simeq (1+dl)\omega$ and the
time $t'=(1-dl)t$ so that the effective action is of the same form as
the bare one but with a renormalized strength of coupling. 

Now we calculate the expression~(\ref{SIeff}) term by term. 

\vskip0.5cm

\underline{\it First order contribution}:

In first order in couplings, it is given by 

\ba
\langle S_I\rangle_f &=& \lambda_0\int dt~\langle {\bf 1}\rangle_f 
- \lambda_1\int dt~\langle\cos\beta [\Phi_{s\Lambda'}(t)+\Phi_f(t)]\rangle_f 
\nonumber \\
& &- \lambda_2\int dt~\langle\cos 2\beta [\Phi_{s\Lambda'}(t)+\Phi_f(t)]
\rangle_f\,.
\label{order1double}
\ea

The expression for correlation function of bosonic exponents reads

\be
\langle\prod_n e^{i\beta_n\Phi(t_n)}\rangle=e^{-\sum_{m>n} \beta_m\beta_n
G(t_m,t_n)}e^{-{1 \over 2}\sum_m \beta_n^2 G(t_m,t_m)} \,.
\label{exp-cor}
\ee
The terms containing Green's functions of coinciding arguments are
singular in the continuous limit. But in our regularized theory they
are finite. The Green's function has the following well known form

\be
G(t_m,t_n) = {1 \over 4\pi} ln\Bigg({R^2 \over {(t_m-t_n)^2+a^2}}\Bigg)\,,
\label{gf}
\ee
where $R$ and $a$ are IR and UV cut-off respectively.
Substituting into (\ref{exp-cor}) we get the result for the
correlation function of bosonic exponents:

\be
\langle\prod_n e^{i\beta_n \Phi(t_n)}\rangle = \prod_{m>n} \Bigg(1+{(t_m-t_n)^2
\over a^2}\Bigg)^{\beta_m\beta_n \over 4\pi} \Bigg({R \over
a}\Bigg)^{-{\big(\sum_n\beta_n\big)^2 \over 4\pi}}\,.
\label{exp-cor1}
\ee
Using the above formulae we obtain

\be
\langle e^{\pm i\beta\phi_f(t)}\rangle_f = e^{-{\beta^2 \over
2}\langle\Phi_f(0)^2\rangle_f} =  
e^{-{\beta^2 \over 2}G_f(0,0)}\,,
\ee
where

\be
-{\beta^2 \over 2}G(0,0) = -{\beta^2 \over 2}\int_0^\Lambda d\omega
f(\omega) = -{\beta^2 \over 4\pi}~ln\Lambda\,,
\label{gengf}
\ee
and

\be
-{\beta^2 \over 2}G_f(0,0) = -{\beta^2 \over
2}\int_{\Lambda-d\Lambda}^\Lambda d\omega f(\omega) = d\Lambda
I'(\Lambda) = -{\beta^2 \over 4\pi}\Big({d\Lambda \over \Lambda}\Big) =
-{\beta^2 \over 4\pi}dl\,.
\label{fastgf}
\ee
Hence (\ref{order1double}) turns out to be 

\ba
\langle S_I\rangle_f &=& \tilde\lambda_0\int \frac{dt}{a}~{\bf 1}
-\tilde\lambda_1\Big(1-\frac{\beta^2}{4\pi}dl\Big) 
\int \frac{dt}{a}~\cos\beta\Phi_{s\Lambda'}(t)  
\nonumber \\
& &- \tilde\lambda_2\Big(1-\frac{4\beta^2}{4\pi}dl\Big) 
\int \frac{dt}{a}~\cos 2\beta\Phi_{s\Lambda'}(t) + O(dl^2)\,,
\label{order1doublefinal}
\ea
where $\tilde\lambda_i = \lambda_i a$ are small dimensionless couplings.

\vskip0.5cm

\underline{\it Second order contribution}:

\vskip0.5cm

We now turn to the quadratic contribution to (\ref{SIeff}). Imposing 
translational invariance it can be written as

\ba
& &{1 \over 2}(\langle S_I\rangle_f^2-\langle S_I^2\rangle_f) 
\nonumber \\
&=&  \frac{\lambda_1^2}{4} \int dt_1 \int dt_2 \Big\{
e^{-\beta^2 G_f(0,0)}
\Big(1-e^{-\beta^2 G_f(t_2-t_2)}\Big)
\cos\beta[\Phi_{s\Lambda'}(t_1)+\Phi_{s\Lambda'}(t_2)]
\nonumber \\
& &+e^{-\beta^2 G_f(0,0)}
\Big(1-e^{\beta^2 G_f(t_2-t_2)}\Big)
\cos\beta[\Phi_{s\Lambda'}(t_1)-\Phi_{s\Lambda'}(t_2)]\Big\}
\nonumber \\
& &+\frac{\lambda_2^2}{4} \int dt_1 \int dt_2 \Big\{
e^{-4\beta^2 G_f(0,0)}
\Big(1-e^{-4\beta^2 G_f(t_2-t_2)}\Big)
\cos 2\beta[\Phi_{s\Lambda'}(t_1)+\Phi_{s\Lambda'}(t_2)]
\nonumber \\
& &+e^{-4\beta^2 G_f(0,0)}
\Big(1-e^{4\beta^2 G_f(t_2-t_2)}\Big)
\cos 2\beta[\Phi_{s\Lambda'}(t_1)-\Phi_{s\Lambda'}(t_2)]\Big\}
\nonumber \\
& &+\frac{\lambda_1\lambda_2}{2}e^{-\frac{5\beta^2}{4\pi}dl}
\int dt_1\int dt_2~\Big\{(1-e^{-2\beta^2G_f(t_1-t_2)})
\cos\beta[\Phi_{s\Lambda'}(t_1)+2\Phi_{s\Lambda'}(t_2)]  
\nonumber \\
& &+\{(1-e^{2\beta^2G_f(t_1-t_2)})
\cos\beta[2\Phi_{s\Lambda'}(t_2)-\Phi_{s\Lambda'}(t_1)]\Big\}\,.
\label{order2double1}
\ea
To evaluate the correlation functions of the
bosonic vertices we have used~(\ref{exp-cor}). 
To evaluate the Green's function for fast moving modes we use the
following scheme. Instead of considering just the momentum shell for
fast moving modes, if we consider all momenta upto cutoff $\Lambda$
then using~(\ref{exp-cor1}) we get

\ba
\langle e^{i\sigma \beta\Phi(t_1)} \cdot e^{i\sigma\beta\Phi(t_2)}\rangle
= \Big(1+{(t_1-t_2)^2
\over a^2}\Big)^{\beta^2 \over 4\pi}\Lambda^{-{\beta^2 \over \pi}}\,,
\ea
which, like~(\ref{gengf}), leads to

\ba
-\beta^2 G(0,0)-\beta^2 G(t_1-t_2) 
= {\beta^2 \over 4\pi}\Big[ln\Big(1+{(t_1-t_2)^2
\over a^2}\Big)-ln\Lambda^4\Big] \, .
\ea
Hence the analogous treatment for fast moving modes follows by performing
the integration in the narrow slice only. Using the expression
for $G_f(0,0)$ given by~(\ref{fastgf}) we arrive at the following
useful expression for fast moving components 

\be
-\beta^2 G_f(t_1-t_2) \approx -{\beta^2 \over
2\pi}dl\Big(ln\Big|\frac{a}{t_1-t_2}\Big|+1\Big)+O(dl^2)\,, 
\label{fastgf1}
\ee
where we have assumed that $|t_1-t_2| \gg a$. It is clear that

\be
G_f(t_1-t_2)=F(r)~dl+O(dl^2)\,,
\ee
where $r = |t_1-t_2|$. If we adopt a sharp momentum cut-off
prescription, $F(r)$ will be Bessel function of order zero,
$F(x)=(1/2\pi)J_0(\Lambda|x|)$, which has a long oscillating tail
and does not fall off rapidly on increasing its argument. However, 
as was shown in~\cite{wieg, kog}, in a
smooth cut-off procedure $F(r)$ is truly short-ranged, essentially
nonzero at $r<\Lambda^{-1} \sim a$. This can be seen in (\ref{order2double1}),
the functions like $1-e^{\pm\beta^2 G_f(t_2-t_2)}$ are also short-ranged. 
This allows us to introduce
the center-of-mass coordinate $\tilde R=(t_1+t_2)/2$ and relative
coordinate $r=t_1-t_2$ and expand the cosines of (\ref{order2double1}) in
$r$. 

The detail calculation is shown in the appendix. Here we only mention 
some of the necessary facts. For example, consider the cosine 
$\cos \beta[\Phi_{s\Lambda'}(t_1)-\Phi_{s\Lambda'}(t_2)] \approx 
1-{\beta^2r^2 \over 2}(\partial_{\tilde R}\Phi_{s\Lambda'}(\tilde
R))^2$, where the first term  
contributes to renormalization of the free energy. In our case that
generates the RG contribution to identity operator. On the other hand,
the second (gradient)$^2$ term is an irrelevant term with a factor
proportional to the UV cut-off $a$ in front of the renormalized
coupling. Here we see a 
striking difference between bulk and boundary RG flows, where in the
latter case the gradient term is responsible for renormalization of
the constant $\beta$. Collecting the $O(\tilde\lambda_1^2)$, 
$O(\tilde\lambda_2^2)$, and $O(\tilde\lambda_1\tilde\lambda_2)$
contributions from the appendix, we arrive at the following 
complete second order expression

\ba
& &\frac{1}{2}(\langle S_I\rangle_f^2-\langle S_I^2\rangle_f)  
\nonumber \\
&=&\tilde\lambda_1^2\frac{\beta^2}{2\pi}dl
\int\frac{dt}{a}~\cos 2\beta\Phi_{s\Lambda'} (t)
-\tilde\lambda_1^2\frac{\beta^2}{2\pi}dl 
\int\frac{dt}{a}~{\bf 1}
\nonumber \\
& &+\tilde\lambda_2^2\frac{2\beta^2}{\pi}dl
\int\frac{dt}{a}~\cos 4\beta\Phi_{s\Lambda'} (t)
-\tilde\lambda_2^2\frac{2\beta^2}{\pi}dl 
\int\frac{dt}{a}~{\bf 1}
\nonumber \\
& &+2\frac{\tilde\lambda_1\tilde\lambda_2}{\pi}\beta^2dl
\int \frac{d\tilde R}{a}~(\cos
3\beta\Phi_{s\Lambda'}(\tilde R) 
-\cos\beta\Phi_{s\Lambda'}(\tilde R))\,. 
\label{order2doublefinal}
\ea

\vskip0.5cm

\underline{\it Third order contribution}:

\vskip0.5cm

The calculation of the cubic contribution to (\ref{SIeff}) is similar
and the details are given in the appendix. Here we give the result
only:

\ba
& &{1 \over
6}\langle S_I^3\rangle_f-{1 \over 2}\langle S_I\rangle_f\langle S_I^2\rangle_f
+{1 \over 3}\langle S_I\rangle_f^3
\nonumber \\
&=&\tilde\lambda_1^3\bigg(\frac{\pi\beta^2}{144}\bigg)dl
\int\frac{dt}{a}~\cos 3\beta\Phi_{s\Lambda'} (t) 
+\tilde\lambda_1^3\bigg(\frac{\pi\beta^2}{144}
\bigg)dl\int\frac{dt}{a}~\cos\beta\Phi_{s\Lambda'} (t)   
\nonumber \\
& &+\tilde\lambda_2^3\bigg(\frac{\pi\beta^2}{36}\bigg)dl
\int\frac{dt}{a}~\cos 6\beta\Phi_{s\Lambda'} (t) 
+\tilde\lambda_2^3\bigg(\frac{\pi\beta^2}{36}
\bigg)dl\int\frac{dt}{a}~\cos 2\beta\Phi_{s\Lambda'} (t)\,. 
\label{order3doublefinal}
\ea

Collecting all the results up to third order from (\ref{order1doublefinal}),
(\ref{order2doublefinal}) and (\ref{order3doublefinal}) we express the
effective boundary action as 

\ba
S_{I~eff}[\Phi_{s\Lambda'}] &=&
\tilde\lambda_0\int\frac{dt}{a}~{\bf 1} 
-\tilde\lambda_1\Big(1-\frac{\beta^2}{4\pi}dl\Big)
\int\frac{dt}{a}~\cos 
\beta \Phi_{s\Lambda'}(t)
-\tilde\lambda_2\Big(1-\frac{\beta^2}{\pi}dl\Big)
\int\frac{dt}{a}~\cos 2\beta \Phi_{s\Lambda'}(t)
\nonumber \\
& &+\tilde\lambda_1^2\frac{\beta^2}{2\pi}dl
\int\frac{dt}{a}~\cos 2\beta\Phi_{s\Lambda'} (t)
-\tilde\lambda_1^2\frac{\beta^2}{2\pi}dl 
\int\frac{dt}{a}~{\bf 1}
\nonumber \\
& &+\tilde\lambda_2^2\frac{2\beta^2}{\pi}dl
\int\frac{dt}{a}~\cos 4\beta\Phi_{s\Lambda'} (t)
-\tilde\lambda_2^2\frac{2\beta^2}{\pi}dl 
\int\frac{dt}{a}~{\bf 1}
\nonumber \\
& &+2\frac{\tilde\lambda_1\tilde\lambda_2}{\pi}\beta^2dl
\int\frac{dt}{a}~\cos
3\beta\Phi_{s\Lambda'}(\tilde R) 
-2\frac{\tilde\lambda_1\tilde\lambda_2}{\pi}\beta^2dl
\int\frac{dt}{a}~
\cos\beta\Phi_{s\Lambda'}(\tilde R)
\nonumber \\
& &\tilde\lambda_1^3\bigg(\frac{\pi\beta^2}{144}\bigg)dl
\int\frac{dt}{a}~\cos 3\beta\Phi_{s\Lambda'} (t) 
+\tilde\lambda_1^3\bigg(\frac{\pi\beta^2}{144}
\bigg)dl\int\frac{dt}{a}~\cos\beta\Phi_{s\Lambda'} (t)   
\nonumber \\
& &+\tilde\lambda_2^3\bigg(\frac{\pi\beta^2}{36}\bigg)dl
\int\frac{dt}{a}~\cos 6\beta\Phi_{s\Lambda'} (t) 
+\tilde\lambda_2^3\bigg(\frac{\pi\beta^2}{36}
\bigg)dl\int\frac{dt}{a}~\cos 2\beta\Phi_{s\Lambda'} (t) \,.
\label{SIeffdouble(semifinal)}
\ea 

In order to restore the original cut-off to $\Lambda$ we rescale the
energy $\omega'=(\Lambda/\Lambda')\omega \simeq (1+dl)\omega$. In
order to keep the product $\omega t$ intact, we have to rescale
time in the opposite way, $t'=(1-dl)t$. The effective action is
of the same form as the bare one but with a renormalized strength of
coupling. Hence neglecting the $O(dl^2)$ terms,

\ba
S_{I~eff}[\Phi_{\Lambda}] &=&
\tilde\lambda_0(1+dl)\int\frac{dt}{a}~{\bf 1} 
-\tilde\lambda_1\Big(1+\Big(1-\frac{\beta^2}{4\pi}\Big)dl\Big)
\int\frac{dt}{a}~\cos\beta \Phi_{s\Lambda'}(t)
\nonumber \\
& &-\tilde\lambda_2\Big(1+\Big(1-\frac{\beta^2}{\pi}\Big)dl\Big)
\int\frac{dt}{a}~\cos 2\beta \Phi_{s\Lambda'}(t)
\nonumber \\
& &+\tilde\lambda_1^2\frac{\beta^2}{2\pi}dl
\int\frac{dt}{a}~\cos 2\beta\Phi_{s\Lambda'} (t)
-\tilde\lambda_1^2\frac{\beta^2}{2\pi}dl 
\int\frac{dt}{a}~{\bf 1}
\nonumber \\
& &+\tilde\lambda_2^2\frac{2\beta^2}{\pi}dl
\int\frac{dt}{a}~\cos 4\beta\Phi_{s\Lambda'} (t)
-\tilde\lambda_2^2\frac{4\beta^2}{2\pi}dl 
\int\frac{dt}{a}~{\bf 1}
\nonumber \\
& &+2\frac{\tilde\lambda_1\tilde\lambda_2}{\pi}\beta^2dl
\int\frac{dt}{a}~\cos
3\beta\Phi_{s\Lambda'}(\tilde R) 
-2\frac{\tilde\lambda_1\tilde\lambda_2}{\pi}\beta^2dl
\int\frac{dt}{a}~
\cos\beta\Phi_{s\Lambda'}(\tilde R)
\nonumber \\
& &+\tilde\lambda_1^3\bigg(\frac{\pi\beta^2}{144}\bigg)dl
\int\frac{dt}{a}~\cos 3\beta\Phi_{s\Lambda'} (t) 
+\tilde\lambda_1^3\bigg(\frac{\pi\beta^2}{144}
\bigg)dl\int\frac{dt}{a}~\cos\beta\Phi_{s\Lambda'} (t)   
\nonumber \\
& &+\tilde\lambda_2^3\bigg(\frac{\pi\beta^2}{36}\bigg)dl
\int\frac{dt}{a}~\cos 6\beta\Phi_{s\Lambda'} (t) 
+\tilde\lambda_2^3\bigg(\frac{\pi\beta^2}{36}
\bigg)dl\int\frac{dt}{a}~\cos 2\beta\Phi_{s\Lambda'} (t) \,.
\nonumber \\
& &
\label{SIeffdouble(final)}
\ea

Note that although initially we considered
perturbation with tachyonic zero, first and second modes, the
non-linear RG flow equations force the coefficients of various other
boundary fields with higher harmonics to evolve. We see from above
expression that the third and fourth harmonics evolve at second order
and the third
and sixth harmonics evolve at third order. Also notice that there is
no mixing between $\lambda_1$ and $\lambda_2$ at third order in the 
coupling. The beta functions can be extracted from the above
expression and are given by   

\ba
\beta_0 &=& \frac{d\tilde\lambda_0}{dl} = \tilde\lambda_0
-\frac{\beta^2}{2\pi}\tilde\lambda_1^2-\frac{2\beta^2}{\pi}\tilde\lambda_2^2
\, ,
\nonumber \\
\beta_1 &=& \frac{d\tilde\lambda_1}{dl} =
-\Big(1-\frac{\beta^2}{4\pi}\Big)\tilde\lambda_1
-\frac{2\beta^2}{\pi}\tilde\lambda_1\tilde\lambda_2
+\frac{\pi\beta^2}{144}\tilde\lambda_1^3 \, , 
\nonumber \\
\beta_2 &=& \frac{d\tilde\lambda_2}{dl} =
-\Big(1-\frac{\beta^2}{\pi}\Big)\tilde\lambda_2
+\frac{\beta^2}{2\pi}\tilde\lambda_1^2
+\frac{\pi\beta^2}{36}\tilde\lambda_2^3 \, .
\label{betafn2}
\ea
      
Now we take $R_{25}=\sqrt 3$, where the perturbation with $\lambda_1$
and $\lambda_2$ becomes least relevant and irrelevant
respectively. Perturbation with the identity operator (associated with
$\lambda_0$) in the RG picture just adds a constant to the action. 
 Typically the
coefficient of the least relevant operator grows at the fastest rate at the
beginning, and drives the system towards the fixed point. 
The final shape of the
soliton is determined by where this fixed point is. 
Rescaling of $\beta:
\beta  \rightarrow \sqrt {4\pi}\beta$ in~(\ref{betafn}) results in

\ba
\beta_0 &=& \frac{d\tilde\lambda_0}{dl} = \tilde\lambda_0
-\frac{2}{3}\tilde\lambda_1^2-\frac{8}{3}\tilde\lambda_2^2
\, ,
\nonumber \\
\beta_1 &=& \frac{d\tilde\lambda_1}{dl} =
-\frac{2}{3}\tilde\lambda_1
-\frac{8}{3}\tilde\lambda_1\tilde\lambda_2
+\frac{\pi^2}{108}\tilde\lambda_1^3 \, , 
\nonumber \\
\beta_2 &=& \frac{d\tilde\lambda_2}{dl} =
\frac{1}{3}\tilde\lambda_2
+\frac{2}{3}\tilde\lambda_1^2
+\frac{\pi^2}{27}\tilde\lambda_2^3 \, .
\label{betafn1}
\ea

Typically single a 
Sine-Gordon boundary perturbation doesn't lead to a nearby fixed point
and the theory typically flows to infinity. Typically $\lambda$
flows all the way from 0 to $\infty$ under renormalization. The
boundary conditions look like Neumann at very high energy (UV) but
like Dirichlet at low energy (IR). The field $\Phi$
satisfy Neumann boundary conditions close to the boundary, but feels
Dirichlet boundary conditions instead far from it. But as in the next
section we will see that Sine-Gordon model with two frequencies 
we considered above has a
nearby fixed point.  

The situation is very analogous to string field theory. Suppose we are
trying to 
construct the lump solution on a circle of radius $R_{25}$. In this case we
need to start with an initial tachyon field for which the coefficient of
$\cos(\Phi/R_{25})$ is non-zero, and then use the equations of
motion of string 
field theory iteratively to improve the solution. This generates all
higher harmonics as well as the tachyon zero momentum mode as we saw in
our RG analysis as well. But if from the
beginning we introduce the tachyon zero momentum mode in the initial
configuration, then the iterative process drives the solution towards the
vacuum solution where the tachyon is constant instead of the one lump
solution. In the RG picture perturbations with the identity operator are
generally regarded as  
uninteresting, as they just add a constant to the action. Also, 
in the peturbative definition via correlation functions, the zero mode
has no effect.


\section{The lump profile and boundary entropy}
\setcounter{equation}{0}

In this section we will calculate the boundary entropy perturbatively
in leading order and compare the result with its exact value.
We follow the method of~\cite{al-prb}.
Let us recall the RG equations~(\ref{betafn2}) obtained in the last section
(up to second order in coupling constants)

\ba
\beta_0 &=& \frac{d\tilde\lambda_0}{dl} = \tilde\lambda_0
-2(1-y_1)\tilde\lambda_1^2-8(1-y_1)\tilde\lambda_2^2
\, ,
\nonumber \\
\beta_1 &=& \frac{d\tilde\lambda_1}{dl} =
-y_1\tilde\lambda_1
-8(1-y_1)\tilde\lambda_1\tilde\lambda_2
\, ,
\nonumber \\
\beta_2 &=& \frac{d\tilde\lambda_2}{dl} =
-y_2\tilde\lambda_2
+2(1-y_1)\tilde\lambda_1^2\,,
\label{betafn}
\ea
where $\Delta_i=1-y_i$ is the scaling dimension of the corresponding
harmonic. The above RG equations satisfies right $\epsilon$-expansion
behaviour according to~\cite{wegner}. 
The first step will be to solve these equations for the bare couplings, 
$\tilde\lambda_0$, $\tilde\lambda_1$, and $\tilde\lambda_2$, as functions 
of renormalized couplings at the scale set by $R$, $\tilde\lambda_0(R)$, 
$\tilde\lambda_1(R)$, and $\tilde\lambda_2(R)$. 

The Pfaffian differential equation obtained from the above set of equations is

\be
[y_1\tilde\lambda_1
+8(1-y_1)\tilde\lambda_1\tilde\lambda_2](d\tilde\lambda_0+d\tilde\lambda_2)
+[\tilde\lambda_0-8(1-y_1)\tilde\lambda_2^2-(4y_1-3)\tilde\lambda_2]d\tilde
\lambda_1 = 0\,,
\label{pfaffian}
\ee 
which is needed in order to satisfy the integrability condition given by

\ba
& &[y_1\tilde\lambda_1+8(1-y_1)\tilde\lambda_1\tilde\lambda_2][24(1-y_1)
\tilde\lambda_2+5y_1-3]+8(1-y_1)\tilde\lambda_1
[\tilde\lambda_0-8(1-y_1)\tilde\lambda_2^2-(4y_1-3)
\tilde\lambda_2]
\nonumber \\
& &+[y_1\tilde\lambda_1+8(1-y_1)\tilde\lambda_1\tilde\lambda_2][1-y_1-8(1-y_1)
\tilde\lambda_2] = 0 \, .
\label{integrability}
\ea
For $\tilde\lambda_1 \neq 0$, we have the following relation between 
$\tilde\lambda_0$ and $\tilde\lambda_2$

\be
64(1-y_1)^2\tilde\lambda_2^2+8(1-y_1)(1+2y_1)\tilde\lambda_2+8(1-y_1)
\tilde\lambda_0-2y_1(1-2y_1)=0\,.
\label{reln1}
\ee
The branch containing the real solution of $\tilde
\lambda_2$ satisfies the following,

\be
8(1-y_1)\tilde\lambda_0 + 2y_1(2y_1-1) \le (2y_1+1)^2/4\,,
\ee 
which implies $\tilde\lambda_0 \leq 0.34$ for $R_{25}=\sqrt 3$,
concistent with our results for the 
desired IR fixed point discussed below. Combining $\beta_0$ 
and $\beta_2$ in~(\ref{betafn}) into

\be
(d\tilde\lambda_0+d\tilde\lambda_2)/[\tilde\lambda_0-8(1-y_1)
\tilde\lambda_2^2-(4y_1-3)\tilde\lambda_2] = dl \, ,
\ee
and using the relation~(\ref{reln1}), we get

\be 
8(1-y_1)d\tilde\lambda_2/[8(1-y_1)\tilde\lambda_2+2y_1-1]=dl\,,
\ee
for $\tilde\lambda_2 \neq \tilde\lambda_2^*$. Integrating the above 
expression from $a$ to $R$, we obtain the following expression for the bare 
coupling $\tilde\lambda_2$ 

\be
\Bigg[\tilde\lambda_2\Big(\frac{R}{a}\Big)^{y_2}\Bigg]=-\tilde\lambda_2^*
\Bigg[\Big(\frac{2y_1-1}{y_1}-\frac{\tilde\lambda_2(R)}{\tilde\lambda_2^*}\Big)
\Big(\frac{R}{a}\Big)
^{y_2-1}-\Big(\frac{2y_1-1}{y_1}\Big)\Big(\frac{R}{a}\Big)^{y_2}\Bigg]\,,
\label{bare2}
\ee
where $\tilde\lambda_2^*=-y_1/8(1-y_1)$ and $1-y_2=\Delta_2=
4y_1-3$ is the scaling dimension of the second tachyon harmonic. 
For $R_{25}=\sqrt 3$, the perturbation with the second harmonic is least 
irrelevant and is added to improve the shape of the profile of the lump 
at IR fixed point. Being irrelavant the perturbing 
operator with $\tilde\lambda_2$
decays out. This can be seen by considering the perturbation expansion
of the two-point correlation function of bosonic exponents. The integral
appearing in the first nonvanishing correction to this correlation 
function converges at large distances for an irrelevant operator. This implies 
that the perturbation expansion does not contain IR singularities; so if
the bare coupling constant is small, its effect will remain small
and will not be amplified in the process of renormalization.  
In the limit the renormalization scale, $R \rightarrow \infty$,

\be
\tilde\lambda_2 = \frac{a}{R}\tilde\lambda_2(R) - \Big(\frac{a}{R}-1\Big)
\Big(\frac{2y_1-1}{y_1}\Big)\tilde\lambda_2^* \rightarrow -0.125 \, ,
\ee  
which is very close to the result of~\cite{msz} for the second harmonic 
at the stable minimum of the tachyon potential. On the other hand, the
other two operators which are relevant for $R_{25}=\sqrt 3$ evolve to the 
values: 
$\tilde\lambda_0^* = 0.25,~\tilde\lambda_1^* \approx -0.35$, 
which are also very close to the results obtained by the level truncation
technique~\cite{msz}. Hence the
lump profile is given by 

\be
T(x_{25})=0.25-0.35\cos (\frac{1}{R_{25}}x_{25})
-0.125\cos (\frac{2}{R_{25}}x_{25})
\ee
as a function of $x_{25}$. The profile
is shown in figure 1. The profile has finite size in contrast to the
result of~\cite{kmm}.

\begin{figure}[htb]
\epsfysize=7cm
\centerline{\epsffile{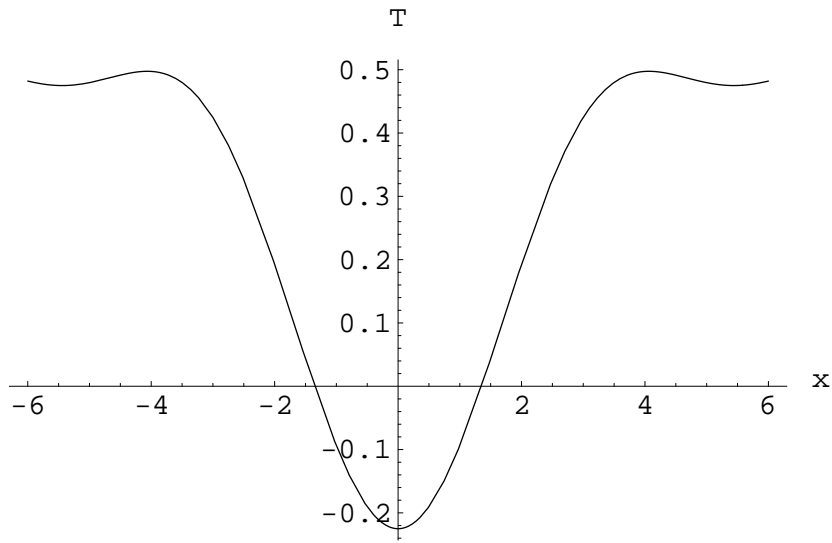}}
\caption{The lump solution with $\lambda_0, \lambda_1, \lambda_2$ 
switched on at $R_{25}=\sqrt 3$.} 
\end{figure} 

Before returning to the calculation of boundary entropy, notice that 
in the RG equations~(\ref{betafn}) in the limit 
$R\to \infty$, the linear terms in $\beta_1$
and $\beta_2$ agree. The reason for these to disagree with $\beta_0$
is that we chose a positive sign in front of $\lambda_0$ in the beginning. 
The RG equations up to linear order are $\frac{d\lambda_i}{dl}
= \pm \Delta_i \lambda_i + O(\lambda_i^2)$, where the sign in front of the
universal linear term depends on the sign of perturbation. RG analysis
with the identity 
operator is trivial and it does not mix with other operators. Also as
we will discuss soon its
contribution goes to ground state energy correction only, not to boundary
entropy. 
  
Next we calculate the bare $\tilde\lambda_0$ and $\tilde\lambda_1$ in
terms of the renormalized couplings. 
In order to obtain the expression for bare coupling, $\tilde\lambda_0$ 
in terms of the renormalized couplings we use~(\ref{reln1}) setting the scale 
at UV cutoff:

\be
\tilde\lambda_0 (a) = -\Big[8(1-y_1)\tilde\lambda_2^2(a)
+(1+2y_1)\tilde\lambda_2(a)+\frac{2y_1(2y_1-1)}{8(1-y_1)}\Big] \, .
\ee
Inserting~(\ref{bare2}) for the bare coupling, $\tilde\lambda_2$, we get the 
following expression for $\tilde\lambda_0$

\be
\Bigg[\Big(\frac{R}{a}\Big)\tilde\lambda_0\Bigg]=-\frac{2y_1-1}{8(1-y_1)}
\Bigg[2y_1-y_1\frac{\tilde\lambda_2(R)}{\tilde\lambda_2^*}
+\frac{a}{R}\Big(1-\frac{2y_1}{2y_1-1}
\frac{\tilde\lambda_2(R)}{\tilde\lambda_2^*}\Big)\Bigg]\,.
\label{bare0}
\ee
Inserting the above expression for bare coupling, $\tilde\lambda_0$ into the 
Pfaffian equation~(\ref{pfaffian}) we get 

\be
\frac{[y_1+8(1-y_1)\tilde\lambda_2]^2d\tilde\lambda_2}
{8(1-y_1)\tilde\lambda_2^2+(3y_1-1)\tilde\lambda_2+y_1(2y_1-1)/[8(1-y_1)]}
=-\frac{d\tilde\lambda_1}{\tilde\lambda_1}\,,
\ee
which, assuming $\tilde\lambda_2 \neq \tilde\lambda_2^*$, simplifies to

\be
\Bigg(\frac{8(1-y_1)+y_1}{8(1-y_1)+2y_1-1}\Bigg)d[8(1-y_1)\tilde\lambda_2]
=-\frac{d\tilde\lambda_1}{\tilde\lambda_1} \, .
\ee
Integrating both sides we obtain

\be
8(1-y_1)\tilde\lambda_2+(1-y_1)~ln[8(1-y_1)\tilde\lambda_2+2y_1-1]
=-ln~\tilde\lambda_1+ln~c\,,
\label{indefinite}
\ee
where the integration constant determines the 
correct trajectory that passes through 
UV and desired IR fixed points. 
One might wonder since many different theories flow to the same fixed
point, that we cannot write the bare couplings in terms of
renormalized couplings. 
In fact, here as we are starting with the correct set of 
RG equations which inherently
have the right UV fixed point as a trivial solution So over an infinitesimal 
segment from the UV fixed point we are automatically on the right
trajectory. What we need to worry about is that as we integrate those
equations we don't deviate from the right trajectory and miss the IR fixed
point. We reach the desired fixed point by determining the correct value of the
integration constant.  
To determine the constant that allows
us to reach the desired {\it nearby} fixed point as the end point 
of the flow we substitute the couplings with their fixed point values 

\be
\tilde\lambda_1^*=-\frac{\sqrt{(3-4y_1)y_1}}{4(1-y_1)}\, ,~~~~~~~~
\tilde\lambda_2^*=-\frac{y_1}{8(1-y_1)}\,,
\ee
into~(\ref{indefinite}) resulting in the following expression for the
bare coupling, $\tilde\lambda_1$ 

\ba
\Bigg[\tilde\lambda_1\Big(\frac{R}{a}\Big)^{y_1}\Bigg]
&=&e^{-(1-y_1)}\Big(\frac{1-y_1}{1-2y_1}\Big)^{1-y_1}\frac{\sqrt{y_1(3-4y_1)}}
{4(1-y_1)}
\Bigg[\Big(\frac{R}{a}\Big)+\Big(\frac{R}{a}\Big)\frac{y_1(1-y_1)}{2y_1-1}
\frac{\tilde\lambda_2(R)}{\tilde\lambda_2^*}
\nonumber \\
& &-(2y_1-1)\Big(1-\frac{y_1}{2y_1-1}\frac{\tilde\lambda_2(R)}{\tilde\lambda_2
^*}\Big)-y_1(1-y_1)\frac{\tilde\lambda_2(R)}{\tilde\lambda_2^*}\Bigg] \, ,
\label{bare1}
\ea
where we have used~(\ref{bare2}).

Having solved the renormalization group equations for the bare couplings as 
functions of renormalized couplings, given by~(\ref{bare0}),~(\ref{bare1}), 
and~(\ref{bare2}) we turn to the calculation for boundary entropy which 
measures the tension of the corresponding D-brane and test the 
$g$-{\it conjecture}, which in turn is related to the minimisation of the
action in the space of open string fields. The boundary has its own free energy
proportional to the size of the system which diverges linearly with scale 
$l=\frac{R}{a}$ in thermodynamic limit. The boundary free energy gets
another contribution independent of scale which counts the boundary degrees
of freedom, measured by $ln~g$,

\be
F_BL=-ln~Z=f_BL-ln~g\,.
\ee   

We first perturbatively expand the partition function $Z$ to 
$O(\tilde\lambda^3)$. We arrive at a UV expansion of the form which
schematically looks like

\be
ln~Z = \sum_n c_n(\tilde\lambda_iR^{y_i})^n
=f_B\Bigg[\tilde\lambda_i\Big(\frac{R}{a}\Big)^{y_i}\Bigg]^{\frac{1}{y_i}}
+ln~g+\Bigg[\tilde\lambda_i\Big(\frac{R}{a}\Big)^{y_i}\Bigg]^{-ve}\,,
\ee 
where $\Delta_i = 1-y_i$ is the scaling dimension. We discard 
terms linear in negative powers of scale and nonuniversal UV 
divergent terms which are linear in scale as they correspond
to ground state energy corrections. The remaining terms have weak
dependence on the scale set by $R$, which acts as an IR cut-off, as we can
absorb them approximately in terms of the renormalized coupling
$\tilde\lambda_i(R)$ 
using the relations~(\ref{bare0}),~(\ref{bare1}), and~(\ref{bare2}) 
obtained by solving simultaneous $\beta$-function equations.
Thus in the $R \rightarrow \infty$ limit (when $\tilde\lambda_i(R) \rightarrow
\tilde\lambda_i^*$), these terms give the contribution to the boundary 
entropy. Note that the weak IR cut-off dependence is consistent with
the renormalization group. 

To cubic order in coupling the partition function is given by

\ba
\frac{Z}{Z_0} &\approx& 1+ \frac{\tilde\lambda_i\tilde\lambda_j}{2!}
\int dt_1dt_2 {\cal T}
\langle \cos\beta_i\Phi(t_1) \cos\beta_j\Phi(t_2) \rangle
\nonumber \\
& &+\frac{\tilde\lambda_i\tilde\lambda_j\tilde\lambda_k}{3!}
\int dt_1dt_2dt_3 {\cal T}
\langle \cos\beta_i\Phi(t_1) \cos\beta_j\Phi(t_2) \cos\beta_k\Phi(t_3)
\rangle\,.
\ea
Since $\langle \cos\beta_i\Phi(t_1) \rangle \propto \Big( \frac{R}{a} \Big)
^{-\Delta_i}$, this vanishes in the thermodynamic limit, $R
\rightarrow \infty$. 
Here $Z_0$ is the partition function at UV fixed point. Any $n$-point
correlation function of Sine-Gordon perturbation, the
electro-nonneutral ($\sum_n\beta_n\neq 0$ in~(\ref{exp-cor1})) part
has an overall factor of $\Big( \frac{R}{a} \Big)^{-\Delta_i}$ that cannot be 
compensated using definition of renormalized coupling $\tilde\lambda_i(R)$; 
therefore in above perturbative expansion contributions come from 
electro-neutral parts only such as $O(\tilde\lambda_1^2)$,
$O(\tilde\lambda_2^2)$, and $O(\tilde\lambda_1^2\tilde\lambda_2)$.
Clearly there is no contribution from the identity operator of any order
as its contribution can be considered to be going to ground state energy 
correction.

The perturbing operator has two point function given by

\be
\langle \cos\beta_i\Phi(t_1) \cos\beta_i\Phi(t_2) \rangle
=\frac{1}{2}\Big(\frac{a}{t_1-t_2}\Big)^{2(1-y_i)}\, .
\ee

So the quadratic contribution to $Z/Z_0$ turns out to be (considering
half-cylinder geometry)

\ba
Z_2 &=& \frac{1}{4} \sum_{i=1,2}
a^{-2y_i} \tilde\lambda_i^2 \int \frac{d\tau_1d\tau_2}
{|\frac{R}{\pi}\sin\frac{\pi}{R}(\tau_1-\tau_2)|^{2(1-y_i)}}
\nonumber \\
&=&\frac{1}{4} \sum_{i=1,2}
a^{-2y_i} \tilde\lambda_i^2 R \int_{-R/2}^{R/2} \frac{d\tau_1}
{|\frac{R}{\pi}\sin\frac{\pi}{R}(\tau)|^{2(1-y_i)}}\,, 
\label{z2a}
\ea
where $\tau=\tau_1-\tau_2$ is the relative coordinate
and $i = 1,2$. The above integration 
is regularized by the cut-off $|\tau| > a$. Mapping the circle to infinite
line: $u=\tan\frac{\pi}{R}\tau$ the above integration becomes

\be
Z_2 = \frac{\pi}{4}\sum_{i=1,2}
\tilde\lambda_i^2\Big(\frac{R}{\pi a}\Big)^{2y_i} 
\int_{|u|<\pi a/R} \frac{du}{(1+u^2)^{y_i}|u|^{2(1-y_i)}}\, .
\label{z2b}
\ee
Integration by parts results in

\ba
Z_2 &=& \frac{\pi}{4}\sum_{i=1,2}
\tilde\lambda_i^2\Bigg(\frac{R}{\pi a}\Bigg)^{2y_i}
2\Bigg[-\frac{1}{(1+u^2)^{y_i}(1-2y_i)u^{1-2y_i}}\Bigg|_{\pi a/R}^{\infty}
-\frac{2y_i}{1-2y_i}\int_0^\infty du~\frac{u^{2y_i}}{(1+u^2)^{1+y_i}}\Bigg]
\nonumber \\
&=& \frac{\pi}{4}\sum_{i=1,2}
\tilde\lambda_i^2\Bigg(\frac{R}{\pi a}\Bigg)^{2y_i}
2\Bigg[\Bigg(\frac{R}{\pi a}\Bigg)^{1-2y_i}
- \frac{2y_i}{1-2y_i}~\frac{1}{2}B(y_i+\frac{1}{2},\frac{1}{2})\Bigg]\, ,
\label{z2c}
\ea
where $B(x,y)$ is the Euler beta function. The first term is
linear in $R$ and does not contribute to boundary entropy.

We will formally perform the computation as an expansion in $y_1$ supposing 
$y_1$ to be small and at the end we will set $y_1 = \frac{2}{3}$
corresponding to $R_{25}=\sqrt 3$. It is not {\it a priori}
obvious that such $y_1$ expansion will converge for finite value of $y_1$.
However, as we will see, it does quite well in our case giving nice
agreement with exact value  
of the ratio $g_{IR}/g_{UV}$.
The situation is similar to computation of critical exponents in three 
dimensions where $\epsilon$-expansion for $4-\epsilon$ dimensions converges
for $\epsilon=1$ and gives nice agreement with the experimental data.
We note that we can reach nearby fixed point at the end of the RG flow 
with $y_1=\frac{2}{3}$ indicating that $y_1$ is small enough.

Now in the small $y_1$ limit $Z_2$ becomes

\be
Z_2 \approx -\frac{\pi^2y_1}{2}\Bigg[\tilde\lambda_1\Big(\frac{R}{\pi a}\Big)
^{y_1}\Bigg]^2\, ,
\label{z2final}
\ee
where we have dropped the $O(\tilde\lambda^2)$ part which vanishes in
the $R \rightarrow \infty$ limit as it has an overall factor of
$\Big(\frac{R}{a}\Big)^{2y_2}$ with $y_2$ negative as can be seen by  
substituting~(\ref{bare2}) in~(\ref{z2c}).

The cubic part is given by

\be
Z_3 = \frac{1}{3!}\frac{\tilde\lambda_1^2\tilde\lambda_2}{4}
\int dt_1dt_2dt_3\langle e^{i\beta\Phi(t_1)}e^{i\beta\Phi(t_2)}
e^{-2i\beta\Phi(t_3)}\rangle
\label{z3a}
\ee  
which on half-cylinder turns out to be

\ba
Z_3 = \frac{1}{3!}\frac{\tilde\lambda_1^2\tilde\lambda_2}{4}
~a^{-2y_1}a^{-y_2}\int d\tau_1d\tau_2d\tau_3
~\frac{\Big|\frac{R}{\pi}\sin\frac{\pi}{R}(\tau_1-\tau_2)\Big|^{2(1-y_1)}}
{\Big|\Big(\frac{R}{a}\Big)^2\sin\frac{\pi}{R}(\tau_2-\tau_3)
~\sin\frac{\pi}{R}(\tau_3-\tau_1)\Big|^{1-y_2}}\, .
\label{z3b}
\ea
Now introducing the relative coordinates $\tilde\tau_1 = \tau_2-\tau_3$
and $\tilde\tau_2 = \tau_1-\tau_3$ and center-of-mass coordinate
$\tilde\tau = (\tau_1 + \tau_2 + \tau_3)/3$ the above integral simplifies to

\ba
Z_3 &=& \frac{1}{3!}\frac{\tilde\lambda_1^2\tilde\lambda_2}{4}
~a^{-2y_1}a^{-y_2}R\int_{|u_i|>\pi a/R} du_1 du_2
~\Bigg(\frac{R}{\pi}\Bigg)^2 \frac{1}{(1+u_1^2)(1+u_2^2)}
\nonumber \\
& &\cdot \Bigg(\frac{R}{\pi}\Bigg)^{2(1-y_1)}
\Bigg(\frac{R}{\pi}\Bigg)^{-2(1-y_2)}
\frac{1}{|u_1/\sqrt{1+u_1^2}|^{1-y_2}|u_2/\sqrt{1+u_2^2}|^{1-y_2}}
\nonumber \\
& &\cdot \Bigg| \Bigg(\frac{u_1-u_2}{1+u_1u_2}\Bigg)\Bigg/\sqrt{1+\Bigg(
\frac{u_1-u_2}{1+u_1u_2}\Bigg)^2}\Bigg|^{2(1-y_1)}\,,
\label{z3c}
\ea 
where $u_i=\tan\frac{\pi}{R}\tilde\tau_i$ with a regularization 
$|u_i|>\pi a/R$, $|u_1-u_2|>\pi a/R$. The above expression can be written in 
a more convenient form given by

\be
Z_3 = \frac{\pi}{4 \times 3!}\Bigg[\tilde\lambda_1\Bigg(\frac{R}{\pi a}\Bigg)
^{y_1}\Bigg]^2\Bigg[\tilde\lambda_2\Bigg(\frac{R}{\pi a}\Bigg)
^{y_2}\Bigg] \int_{|u_i|>\pi a/R} du_1 du_2~
\frac{|u_1-u_2|^{2(1-y_1)}}{[(1+u_1^2)(1+u_2^2)]^{y_1}(|u_1|
|u_2|)^{1-y_2}}\, .
\label{z3d}
\ee
Let us first concentrate on terms of $O[(R/a)^{3y_1}]$  and the ground state 
energy contribution from terms of $O(R/a)$. To achieve that we make a change
of variables: $u_2 = vu_1$. 
The above integral becomes

\be
Z_3 = \frac{\pi}{4 \times 3!}\Bigg[\tilde\lambda_1\Bigg(\frac{R}{\pi a}\Bigg)
^{y_1}\Bigg]^2\Bigg[\tilde\lambda_2\Bigg(\frac{R}{\pi a}\Bigg)
^{y_2}\Bigg]\int_{|u_1|>u_0(v)}du_1dv~\frac{(1-v)^{2(1-y_1)}}
{|v|^{4(1-y_1)}|u_1|^{5-6y_1}[(1+u_1^2)(1+u_1^2v^2)]^{y_1}}\, .
\label{z3e}
\ee
where $u_0(v) \equiv \mbox {max}\{\frac{\pi a}{R},
\frac{\pi a}{R|v|},\frac{\pi a}
{R|1-v|}\}$ is the new cut-off. Integration by parts with respect to $u_1$
gives

\ba
Z_3 &=& \frac{\pi}{4 \times 3!}\Bigg[\tilde\lambda_1\Bigg(\frac{R}{\pi a}\Bigg)
^{y_1}\Bigg]^2\Bigg[\tilde\lambda_2\Bigg(\frac{R}{\pi a}\Bigg)
^{y_2}\Bigg]\int_{-\infty}^{\infty}dv
\nonumber \\
& &\cdot\frac{|1-v|^{2(1-y_1)}}{|v|^{4(1-y_1)}}
\Bigg[\frac{-2}{[(1+u_1^2)(1+v^2u_1^2)]^{y_1}(4-6y_1)u_1^{4-6y_1}}\Bigg|
_{u_0(v)}^\infty
\nonumber \\
& &-\frac{2y_1}{4-6y_1}\int_{-\infty}^{\infty}
\frac{du_1}{[(1+u_1^2)(1+v^2u_1^2)]^{y_1}|u_1|^{5-6y_1}}
\Bigg(\frac{u_1^2}{1+u_1^2}+\frac{v^2u_1^2}{1+v^2u_1^2}\Bigg)\Bigg]\,.
\label{z3f}
\ea
From the surface term we will get contribution 
of $O\Bigg[\Bigg(\frac{R}{\pi a}\Bigg)\Bigg]$ 
which has no effect on boundary entropy. 
The rest is UV finite and we can remove the cut-off. However there will be a 
contribution of $O\Bigg[\Bigg(\frac{R}{a}\Bigg)^{2y_1}\Bigg]$ coming from 
$O\Bigg[\Bigg(\frac{a}{R}\Bigg)^{y_1}\Bigg]$ term in the denominator. 
To extract this contribution we perform the following calculation.  

We first introduce step functions into the integral~(\ref{z3d})

\be
I(\epsilon) = \int_{-\infty}^{\infty}du_1du_2~\frac{|u_1-u_2|^{2(1-y_1)}}
{[(1+u_1^2)(1+u_2^2)]^{y_1}(|u_1|
|u_2|)^{4(1-y_1)}}\theta(u_1^2-\epsilon^2)\theta(u_2^2-\epsilon^2)
\theta[(u_1-u_2)^2-\epsilon^2]\, ,
\ee
where $\epsilon \equiv a\pi/R$ is the cut-off. It is convenient
to differentiate the integral with respect to $\epsilon$:

\be
\frac{dI(\epsilon)}{d\epsilon}
\approx \frac{6}{|\epsilon|^{4(1-y_1)}}\int_{-\infty}^\infty 
\frac{du_1}{(1+u_1^2)^{y_i}|u_1|^{2(1-y_i)}}~\theta(u_1^2-\epsilon^2)\, .
\ee
The integral is identical to~(\ref{z2b}) with the same cut-off. 
So using previous result obtained for $Z_2$ we get

\be
\frac{dI(\epsilon)}{d\epsilon} = \frac{6}{\epsilon^{4(1-y_1)}}\times
2[\epsilon^{-(1-2y_1)}-\pi y_1]\,,
\ee
that gives 

\be
I(\epsilon) = \frac{12}{6y_1-4}\Bigg(\frac{R}{\pi a}\Bigg)^{4-6y_1}
-\frac{12\pi y_1}{y_2}\Bigg(\frac{R}{\pi a}\Bigg)^{-y_2} + \hbox{const.}
\label{I}
\ee
The second part gives the desired 
$O\Bigg[\Bigg(\frac{R}{a}\Bigg)^{2y_1}\Bigg]$ 
behaviour that has weak cut-off dependence. But the first part is of no
interest to us since it is the ground state energy contribution from $Z_3$.
The constant part comes from evaluating the integral~(\ref{z3f})
throwing away the
surface term. Writing the integral in terms of old variables $u_1,u_2$ we get

\ba
&&-\frac{2\pi y_1}{4 \times 3!(4-6y_1)}
\Bigg[\tilde\lambda_1\Bigg(\frac{R}{\pi a}\Bigg)
^{y_1}\Bigg]^2\Bigg[\tilde\lambda_2\Bigg(\frac{R}{\pi a}\Bigg)
^{y_2}\Bigg]
\nonumber \\
& &\cdot\int_{-\infty}^{\infty}\frac{du_1du_2~|u_1-u_2|^{2(1-y_1)}}
{[(1+u_1^2)(1+u_2^2)]^{y_1}|u_1u_2|^{4(1-y_1)}}
\Bigg(\frac{u_1^2}{1+u_1^2}
+\frac{u_2^2}{1+u_2^2}\Bigg) 
\nonumber \\
&&=-\frac{\pi y_1}{3!(4-6y_1)}
\Bigg[\tilde\lambda_1\Bigg(\frac{R}{\pi a}\Bigg)
^{y_1}\Bigg]^2\Bigg[\tilde\lambda_2\Bigg(\frac{R}{\pi a}\Bigg)
^{y_2}\Bigg]\int_{-\infty}^{\infty}\frac{du_1du_2~|u_1-u_2|^{2(1-y_1)}}
{[(1+u_1^2)(1+u_2^2)]^{y_1}|u_1u_2|^{4(1-y_1)}}\Bigg(\frac{u_1^2}{1+u_1^2}
\Bigg)\,.
\nonumber \\
& & 
\ea
We notice that in small $y_1$ limit, while performing the $u_2$ integration, 
the $y_1$ factor in front is cancelled by a divergence from $u_2\approx 0$:

\be
2\int_{\pi a/R}^\infty\frac{du_2~(u_1-u_2)^{2(1-y_1)}}{(1+u_2^2)^{y_1}
|u_2|^{4(1-y_1)}}
\approx -\frac{2|u_1|^{2(1-y_1)}}{4y_1-3}~\Bigg(\frac{\pi a}{R}\Bigg)
^{4y_1-3}\, .
\ee
Hence in small $y_1$ limit, the constant part of~(\ref{I}) becomes

\ba
&&-\frac{2\pi y_1}{3!(4-6y_1)(3-4y_1)}
\Bigg[\tilde\lambda_1\Bigg(\frac{R}{\pi a}\Bigg)
^{y_1}\Bigg]^2\Bigg[\tilde\lambda_2\Bigg(\frac{R}{\pi a}\Bigg)
^{y_2}\Bigg]\Bigg(\frac{\pi a}{R}\Bigg)^{4y_1-3}\int_{-\infty}^\infty
\frac{du_1}{(1+u_1^2)^{y_1}|u_1|^{2(1-y_1)}}\Bigg(\frac{u_1^2}{1+u_1^2}\Bigg)
\nonumber \\
&&\approx -\frac{\pi^2y_1}{3!\times 6}
\Bigg[\tilde\lambda_1\Bigg(\frac{R}{\pi a}\Bigg)
^{y_1}\Bigg]^2\Bigg[\tilde\lambda_2\Bigg(\frac{R}{\pi a}\Bigg)
^{y_2}\Bigg]\Bigg(\frac{R}{\pi a}\Bigg)^{-y_2}\, .
\ea

So ignoring higher orders in $y_1$ 
the final expression for $Z/Z_0$ contributing to boundary entropy is
given by
\be
\frac{Z}{Z_0} = 1-\frac{\pi^2}{2}y_1\Bigg[\tilde\lambda_1\Bigg(\frac{R}{\pi a}
\Bigg)^{y_1}\Bigg]^2
-\frac{\pi^2y_1}{3!}\Bigg[\frac{1}{6}-\Bigg(1+\frac{4y_1}{3}\Bigg)\Bigg]
\Bigg[\tilde\lambda_1\Bigg(\frac{R}{\pi a}\Bigg)
^{y_1}\Bigg]^2\Bigg[\tilde\lambda_2\Bigg(\frac{R}{\pi a}\Bigg)
^{y_2}\Bigg]\Bigg(\frac{R}{\pi a}\Bigg)^{-y_2}\,.
\ee
Having already obtained the expression for the bare couplings 
in terms of the renormalized couplings 
as given by~(\ref{bare2}) and~(\ref{bare1}) we can rewrite the above
result in terms of renormalized couplings only. 
As $R \rightarrow \infty$, $\tilde\lambda_i(R) \rightarrow \tilde\lambda_i^*$.
Hence the desired ratio between ground state entropy at IR and UV fixed points
turns out to be (in the leading order in $y_1$)

\be
r_p\equiv \frac{g_{\mbox{IR}}}{g_{\mbox{UV}}} = \lim_{R \rightarrow 
\infty} 
\frac{Z}{Z_0} \approx 1-0.50088964~y_1^2 \, .
\ee
Notice that the change in $g$ is negative, implying decrease of $g$ under flow 
between UV to IR point. Also the ratio becomes unity for exactly marginal
case, {\it i.e.} where $y_1 = 0$, giving a line of fixed points.
On the other hand the exact result is given by~\cite{fsw-exact} 

\be
r_e\equiv \frac{g_{\mbox{IR}}}{g_{\mbox{UV}}} = \sqrt{1-y_1}\, .
\ee
For $R_{25}= \sqrt 3$, perturbative result is $r_p \approx 0.65695$
compared to the exact result, $r_e \approx 0.57735$; so 
our result is within $13\%$.
On the other hand for $R_{25}= 1.1$, {\it i.e.} when $y_1$ is very
small and the perturbation by the first tachyon harmonic is nearly marginal,
the perturbative result is $r_p \approx 0.98491$ compared to
the exact result, $r_e \approx 0.90909$; the perturbative result is
now obtained to an accuracy of $8\%$ of the exact result.

\newpage

\section{Multicriticality, $U(\infty )$ and Kondo picture of IR fixed point}

In the effective Landau-Ginzburg description of two dimensional
conformal field theory,

\be
{\cal L} = \int d^2x~(\frac{1}{2}(\partial\Phi )^2+V(\Phi))\, ,
\label{LG}
\ee
$\Phi$ being the order parameter for some physical system, the 
extrema of the general polynomial interaction correspond to
various critical phases of the system. Many systems possess a 
$\Phi \rightarrow -\Phi$ symmetry with an even polynomial interaction 
$V(\Phi )=\sum_m g_m\Phi^{2m}$.
For a polynomial $V(\Phi )$ of degree
$2(m-1)$, this ensures existence of $(m-1)$ minima separated by $(m-2)$
maxima. Several critical phases can coexist if the corresponding 
extrema coincide. Hence the most critical potential is a monomial in $\Phi$ and
the $(m-1)$ critical behaviour of the theory is given by the interaction:

\be
{\cal L}_{int} = g\int d^2x~\Phi^{2(m-1)}\,.
\label{mono}
\ee
By comparing the structure of the operator algebra of the above bulk 
critical theory with that of the unitary diagonal minimal model $M_{m+1,m}$,
characterized by central charge $c(m)=1-\frac{6}{m(m+1)}$ for $m=3,4,\ldots ,$ 
Zamolodchikov~\cite{zamo-multi} has shown that each $(m-1)$ multicritical
behaviour of the theory~(\ref{LG}) is nothing but a minimal model, $M_{m+1,m}$.

Generalizing this concept to boundary conformal field theory we consider
the boundary potential of the form

\be
V(\Phi ) = \sum_k g_k \Phi^{2k}(0)\,.
\ee
The bulk theory is still the $c=1$ ({\it i.e.} 
$M_{\infty +1,\infty}$) minimal model throughout the flow. Near the UV
fixed point, as we turn on the boundary interaction (which can be 
considered as even polynomial interaction of order infinity and hence an
indication of underlying infinite number of critical phases of the system) 
the system flows to the IR near which the system has finite
multicritical 
behaviour. The strength of multicriticality ({\it i.e.} how many critical
phases can coexist) depends on number of couplings
involved, {\it i.e.} on dimensionality of the space of coupling constants.
In other words, expanding the Sine-Gordon polynomial about $X_{25}(0,\tau )
=0$ near the IR fixed point, the potential has the following form:

\be
V(\Phi ) = (\sum_n \tilde\lambda_n^*)
-(\sum_n \tilde\lambda_n^*\frac{n^2}{2!R^2})\Phi (0)^2 
+(\sum_n \tilde\lambda_n^*\frac{n^4}{4!R^4})\Phi (0)^4-\ldots 
\equiv (-1)^k\sum_k g_{k}^*\Phi (0)^{2k}\, . 
\ee
Now using the results from previous section, 
$g_0^* = \sum_n (-1)^n\tilde\lambda_n^* \approx 0.25-0.3535+0.25-\ldots
=0.146-\ldots\sim 0$, so 
one can consider $g_0^*\rightarrow 0$ for a large number  
of couplings turned on. 
For the perturbation with the first three tachyon
harmonics~(\ref{double-sine-gordon}), 
{\it i.e.} $n\leq 2$, we observe that for $R_{25}>1$, only
$g_1^*\sim 0.108 \neq 0,~g_2^*\sim 0.051\neq 0$, and all other $g_i^*\sim
0.0008\sim 0$ indicating $V(\Phi )\sim g_0^*+g_1^*\Phi^2+g_2^*\Phi^4$ 
and hence the existence of two critical phases of the system at the IR
fixed point.  
Also for $R_{25}$ close to one ({\it i.e.} the self-dual radius), 
$g_1^*\sim 0.325$, $g_2^*
\sim 0.152\neq 0$ showing the two phases again. 
As we turn on more and more couplings, we will see that lower order
coefficients of  
$V(\Phi )$ become zero, but some higher order coefficient becomes nonzero 
producing a shift in the degree of the polynomial effective interaction 
indicating the presence of more and more critical phases. For $n
\rightarrow \infty$ the 
effective interaction is basically a monomial of very large order. 
We can expect that by turning on 
all possible couplings ({\it i.e.} infinite number of tachyon modes) we can
probe the $\infty$-multicritical behaviour of the system. 

Note that the values
of coupling constants $g_k^*$ are meaningful only in the context of a
particular 
renormalization scheme and when the composite fields $\Phi^{2k}$ are defined.
However the multicritical behaviour observed is RG scheme 
independent~\cite{zamo-multi}.
For the effective polynomial interaction $V(\Phi ) = 
\sum_{k=1}^Ng_k\Phi^{2k}(0)$, the $m$-critical behaviour ({\it i.e.} presence
of $m$ degenerate minima) of the potential is confined to hypersurfaces
$S_m$, $m=2,3,4,\ldots ,N$ of codimension $m-2$, {\it i.e.} of dimension
$N+2-m$. The $N+1-m$ dimensional boundary $C_m$ of this hypersurface 
$S_m$ is critical. Whereas the form of the hypersurfaces $S_m$ and $C_m$ 
essentially depends on the regularization scheme of the theory, the 
$m$-critical behaviour on the hypersurface $C_m$ is universal and depends 
only on $m$. For the Sine-Gordon potential (essentially an infinite degree
polynomial interaction), $m=2,3,4,\ldots,\infty$-critical behaviours are
possible and our observation is that by turning on full set of coupling
constants we can fully explore the $\infty$-critical behaviour at IR.

In string theory language, the $m$-critical behaviour 
in the IR is nothing but a free $c=1$ boundary
conformal field theory of $m$-coalescing D24-branes with $U(m)$ 
Chan-Paton factor. Note that in order to 
probe the $\infty$-critical behaviour (in other words $U(\infty )$ symmetry
of the nonperturbative bosonic vacuum reached asymptotically from the
core of the D24-brane) we need to turn on all tachyonic modes even though
higher modes, being highly irrelevant, have no significant 
effect in perturbative RG analysis. When we perturb by the zeroth mode only 
(where $m=0$) we end up with the closed string
vacuum with no D-brane ({\it i.e.} $g_{IR}/g_{UV}\rightarrow 0$).

At this point we can draw an interesting analogy from Kondo
physics\footnote{For a treatment on CFT description of it
see~\cite{affleck-cftkondo}. For a recent discussion on Kondo physics in
the context of tachyon condensation in open string field theory
see~\cite{bk}.}. The Kondo model describes the interaction between
$k$-degenerate  
bands of spin-$\frac{1}{2}$ conduction electrons and a quantum
impurity of arbitrary size $s$ placed at one boundary, say at origin,
which essentially makes the system nontranslationally invariant.
This type of problem is similar to the open string theory in the sense that 
boundary spin is like a dynamic degrees of freedom or Chan-Paton factor.  
The IR (or low-temperature) behaviour is quite different depending on
the relative size of $2s$ and $k$.  
For the {\it underscreened case} 
({\it i.e.} $s>\frac{k}{2}$) the RG flow leads to 
IR fixed point with Fermi liquid behaviour. In the course of RG flow the
boundary spin fuses with internal spin of the system and the electron
channel screens or swallows up a part of the impurity spin giving rise
to a similar current algebra as that of the UV fixed point with a shifted 
Kac-Moody current, 

\be
\overrightarrow{\cal J}_n=\overrightarrow{J}_n+\overrightarrow{S}\,,
\ee
leading to rearrangement in the conformal towers of the theory
\footnote{For exactly screened and underscreened cases, $s \geq \frac{k}{2}$,
the IR fixed point is given by fusion with spin-$\frac{k}{2}$ priamry. The
fusion rules are~\cite{al-fusion} $j \otimes \frac{k}{2} =
\frac{k}{2}-j$. Each conformal 
tower is mapped into a unique conformal tower giving rise to free fermion
spectrum with a $\pi/2$ phase shift. Whereas IR fixed point in overscreened
case is given by fusion with spin-s primary.}. 
The fact that the IR fixed point for the underscreened case is
stable can be seen from the interaction between the partially screened
spin and the electrons on the neighbouring lattice sites, which is
ferromagnetic and hence irrelevant. The strong coupling fixed point is
much the same as the UV or zero coupling fixed point. Only the size of
the impurity spin is reduced and there is a change in boundary
condition on the otherwise free fermions (corresponding to a $\pi/2$
phase shift). This is precisely what happens in the present context of
tachyon condensation which leads to a change of boundary condition from
Neumann to Dirichlet. The leftover impurity 
spin $s-\frac{k}{2}$ at IR decouples from the internal
symmetry group and is similar to the $U(k)$ Chan-Paton factor of $k$ D24-branes
at IR. 

For {\it exact screening}, $s=\frac{k}{2}$, the internal spin fully
swallows up the boundary spin giving rise to a translationally invariant 
system which is analogous to translationally invariant nonperturbative  
closed string vacuum with no D-brane. 

For the {\it overscreened} case ({\it i.e.} $s<\frac{k}{2}$), the IR
fixed point is very nontrivial and is determined by fusion with the 
spin-$s$ primary.   
The theory arrives at non-Fermi liquid fixed point in the IR
which describes completely different physics. The strong coupling IR
fixed point ground state has an overscreened spin of size
$\frac{k}{2}-s$. The induced interaction is antiferromagnetic and
hence unstable. The existence of such nontrivial fixed point can be proved
in the large-$k$ limit (keeping $s$ fixed). The ground state cannot be
described by a simple 
physical picture. In~\cite{al-overscreen} the overscreened case is
described based on hamiltonian non-abelian bosonization. The idea is
to represent the $2k$ species of electrons in terms of three
independent bosonic fields: a free scalar carrying $U(1)$ charge, a
$SU_k(2)$ WZW non-linear $\sigma$-model field carring $SU(2)$ spin, and
a $SU_2(k)$ WZW field carring $SU(k)$ flavour degrees of freedom.
The Kondo interaction involves only the spin current. The degrees of
freedom at the IR fixed point are described by an $U(1)\times
SU_k(2)\times SU_2(k)$ Kac-Moody invariant CFT. Both boundary scaling
dimensions and Fermi-liquid bulk scaling dimensions are described by a
$U(1)\times SU_k(2)\times SU_2(k)$ Kac-Moody boundary CFT. 
There are restrictions on the combinations of charge, spin, flavour
degrees of freedom. Unlike the underscreened case, here fermion
exponents are not recovered and are generally not half-integer.
They come from more
general combinations of three types of bosonic fields. The fermion
exponents are sums of $U(1)$, $SU_2(k)$ and $SU_k(2)$ Kac-Moody
current algebra exponents. They are not free and in fact {\it bound}
together in combinations to form fermion composites. 

The analogue of overscreened Kondo physics in string theory is the case
when one considers string theory on the $SU(2)$ group manifold that can be
described by a world-sheet WZW action and D-branes are stabilized
against shrinking by quantized $U(1)$ flux as discussed by 
Bachas, Douglas and Schweigert~\cite{bds} (also see~\cite{ars}). They
considered a static D2-brane wrapping an $S^2$ parametrized by $(\theta
,\phi)$ breaking $SU(2)_L\times SU(2)_R$ to the diagonal $SU(2)_{diag}$. 
With non-zero background $B$ field the gauge invariant quantity is ${\cal
F} = B+2\pi F$. If $0<n<k$, where $n$ is the magnetic monopole number
appearing due to the world-volume flux quantization $\int F$, the D2
brane is prevented from shrinking to one of the poles of
$S^3$~\cite{bds}. Inside this range the energy of the system has a
unique minimum away from the poles of $S^3$. In the large-$k$ limit,
that minimum energy reduces to the mass of $n$ D-particles. The stable
configuration leads in the dual picture to a bound state of $n$
D-particles on $S^2$ similar to the case described in~\cite{myers}.
 
From the above physical picture (of the underscreening and exact screening
Kondo effect) one can try to see the mechanism of 
generation of local $U(m)$
Chan-Paton factors of $m$ coincident D24-branes as a result of breaking of 
inherent stringy $U(\infty )$ symmetry due to fusion of boundary spin 
or dynamical degrees of freedom with
internal symmetry. The local operator ${\cal O}=g_m\Phi^{2m}(0)$ near
the IR
fixed point breaks $U(\infty )$ down to $U(\infty -m)
\times U(m)$ in the core of the soliton. While there is a local $U(m)$
on $m$ 
coincident D24-branes, $U(\infty -m)$ is again swallowed up by the internal 
symmetry of the system. As $m\rightarrow 0$ (the exact
screening case), the situation is similar to perturbation
by the tachyon zero mode only, which leads the theory to the nonperturbative
closed string vacuum and the local $U(1)$ symmetry on the D25-brane is
restored to the full $U(\infty )$ symmetry. For $m \rightarrow
\infty$ (when 
perturbations with all $\lambda_i$s are triggered) we get full
$U(\infty )$ symmetry 
corresponding to infinite number of D24-branes in the core of the
soliton. Symmetry restoration in the same context for noncommutative
tachyon solitons is discussed in~\cite{gms}.

It can also be seen whether the operator ${\cal O}=g_m\Phi^{2m}(0)$ can 
give rise to $m^2$ copies of identity operator at IR fixed point,
which are basically $m^2$ generators of $U(m)$ as all operators 
flow to identity operators or derivative of identity operators at IR
fixed point. So from the multicritical behaviour discussed above, four
copies of the identity operator are left, indicating the final
configuration to be that of two D24-branes.


\section{Final remarks}

There are many unexplored questions related to the material covered in
this paper. We mention some of them below along with some remarks.

It would be interesting to understand the precise relationship between
boundary string field theory and Chern-Simons string field theory. 
In other words, how the action in boundary SFT is related to the 
Chern-Simons action in cubic SFT. It is pointed out in~\cite{kmm} that
the boundary SFT action described in certain coordinates on the space of
coupling constants must be related to the cubic action described in
a particular choice of coordinates on the space of string fields. The two
choices of coordinates are related by some complicated singular transformation.
In~\cite{kmm} the boundary conformal field theory is perturbed by a 
simple tachyon profile with mass parameter $u$ that flows from zero
in the UV to infinity in the IR. The lump profile has width $1/u$ which 
vanishes in the IR. In contrast, the level truncation scheme in cubic
SFT gives solitons with finite width\footnote{This in fact is not a 
disagreement since in the corresponding Zamolodchikov metric on the 
field space the distance between the perturbative UV fixed point and the 
stable minimum is finite.}. 

On the other hand, in our scheme one gets the soliton profile
with finite width situated in a nearby IR fixed point. The set of 
values the first three tachyon harmonics evolve to in order to hit the 
desired IR fixed point is very close to the one obtained in the level
truncation  
scheme in cubic SFT~\cite{msz}. Also the perturbative result of the boundary
entropy is in good agreement with the exact result. It might be more
appropriate to choose the signs of all the couplings initially to be
the same or by a symmetry to be of the same sign for the zero and
second modes and of 
opposite sign for the first mode. The former choice does not make the
analysis different - we have already mentioned that the zero mode is the
identity operator which appears in the RG equations only in linear
order and does not contribute to the boundary entropy (which is the
only physical quantity of interest here). The latter choice (that
involves opposite signs for first and second modes) makes the
calculation of the boundary entropy more complicated and it does not
fit in well
with the setup used in~\cite{al-prb}. While the relationship between
the our setup and the Chern-Simons SFT is not clear, finding the reason
for this (apparent) agreement of 
our result with that of~\cite{msz} is left for future work.  

There is a striking similarity between the beta function
equations~(\ref{betafn2}) in the first quantized theory and the string
field equations of motion given below (in the notation of~\cite{msz}):

\ba
2\frac{\partial V}{\partial t_0} &=&
2t_0-2K^3t_0^2-K^{3-\frac{2}{R^2}}t_1^2-K^{3-\frac{2.2^2}{R^2}}t_2^2=
0\, , 
\nonumber \\
2\frac{\partial V}{\partial t_1} &=&-\Big(1-\frac{1}{R^2}\Big)t_1
+2K^{3-\frac{2}{R^2}}t_0t_1 + K^{3-\frac{2.3}{R^2}}t_1t_2=0
\, ,
\nonumber \\
2\frac{\partial V}{\partial t_2} &=&-\Big(1-\frac{4}{R^2}\Big)t_2
+2K^{3-\frac{2.4}{R^2}}t_0t_2 + K^{3-\frac{2.3}{R^2}}t_1^2 =0\, ,
\label{sfteqn}
\ea
where $K=\frac{3\sqrt 3}{4}$ is the inverse of the mapping radius of
the punctured disks defining the three string vertex. It is maximal
for the vertex of the cubic potential~\cite{bel-zw}. 
The main difference is in the zero mode dependence. In the RG case the
identity operator adds a constant to the action. In the RG equation
the corresponding
coupling constant appears in the universal linear order only. Presumably there 
are definite complicated nonlinear relations between the different
tachyon harmonics generating term proportional to $t_0^2$ and the cross terms
$t_0t_1$, $t_0t_2$ in RG equations. On the 
other hand since $\tilde\lambda = a\lambda$ is dimensionless, 
the mapping radius $K^{-1}$ should be somehow related to the cut-off $a \sim
\Lambda^{-1}$ in boundary conformal field theory side.

In order to get a concrete relationship between our analysis with cubic
string field theory we do need to
develop a concrete proposal for how we can define an off-shell string
field theory using our variables; this requires choosing an appropriate
UV regulator that makes the equations of motion for all the coupling
constants unambiguous and free of divergence. Once this is done, one can
then interprete the fixed points that we have found as (approximate) 
solutions of the equations of motion of this specific string field
theory. In the absence of such a field theory it is not clear what these
solutions 
correspond to. These are of course the RG fixed points in a specific
renormalization scheme, but this is a first quantized viewpoint as opposed
to a second quantized one. Cubic SFT clearly provides such a formulation
- the equations of motion involving arbitrarily high level fields do not
suffer from UV divergence.  The background independent SFT does not
provide such a formulation for all fields, but can be solved exactly for a
special class of configurations (corresponding to operators quadratic in
$X$) and hence we can consider a subsector of the theory where only these
operators are switched on. In this way it avoids having to do with
non-renormalizable field theories. In your formulation
we have switched a finite subset of the operators in an approximation scheme,
but this in higher order would induce operators of arbitrarily high
dimensions and so one needs to understand how to deal with
non-renormalizable field theories. One way to avoid this might be to use
the techniques of integrable perturbations developed by Ghoshal and
Zamolodchikov~\cite{gz}.

We have not raised issues concerning the shape and size of the lump
representing the lower dimensional D-brane. As was pointed out
in~\cite{msz}, in order to get some insight into this issue one needs to find
the energy density of the object. Since cubic SFT is nonlocal, it is
presently not clear how to address this. 
Also the size might be an artifact of the particular gauge chosen.
The shape and size depends on a particular definition of off-shell
string field, and can vary from one SFT to another. 
The important fact is that in Chern-Simons SFT everything is smooth and 
non-singular, so it gives a good definition of the off-shell string field
\footnote{On the other hand due to UV divergences on the worldsheet
it is difficult to define a general off-shell amplitude in boundary
SFT and one needs to follow an appropriate regularization scheme.}.

There is no invariant definition of size of the D-brane unless one 
looks at the coupling 
to metric or some closed string background. Presumably one needs to 
follow the analysis analogous to the bulk conformal perturbation theory
in the presence of a fluctuating metric (see for example~\cite{kut,hsu-kut}). 
It is pointed out in~\cite{zamo-82}
that the dynamics of the metric makes sense only when there is a scale 
in the path integral, for example by fixing the area. This is done by
the so-called {\it gravitational dressing} of the microscopic matter
operator by the appropriate Liouville dependence and inserting it in the
path integral. An analogous treatment in boundary perturbation theory
might give some useful insight regarding the shape and size of the
D-brane. 

An other important issue is to extend the scheme used in this paper to
superstring theories where lower dimensional D-branes are identified 
with a tachyonic kink solution rather than a lump solution\footnote{Very
recently superstring generalization of the results of~\cite{kmm} has
appeared~\cite{kmm1}.}. In the 
supersymmetric setup it is more convenient and sensible to study a 
configuration of parallel D-branes. Also the size and shape of the
D-branes in the IR can be investigated considering physical picture
involving closed string scattering off the D-branes~\cite{dkps}. It
would also be of interest to study intersecting D-brane configurations
using the boundary RG analysis discussed in this paper.


\section*{Acknowledgements}

We are grateful to Michael Douglas and Michael Green for many
interesting discussions and critical reading of the manuscript. We
thank David Kutasov and Ashoke Sen for very useful discussions,
correspondence and important remarks 
on a preliminary draft. We thank Alexander Zamolodchikov      
for a fruitful discussion on boundary entropy. SD is grateful to DAMTP
for its hospitality in the early stages of this work. TD is grateful
to Rutgers New High Energy Theory Center for its hospitality in the final
stages of this work. We acknowledge partial support from the PPARC SPG
programme, ``String Theory and Realistic Field Theory'',
PPA/G/S/1998/00613. TD is supported by a Scholarship of Trinity
College, Cambridge.


\appendix
\section{Appendix} \label{ap:a}
\setcounter{equation}{0}

In this appendix we will present the details of the RG calculation in order
two and three.

\vskip0.5cm

\underline{\bf Second order contribution:}

\vskip0.5cm

Using our ansatz for Green's function of fast moving modes~(\ref{fastgf1}),
the $O(\lambda_1^2)$ part of~(\ref{order2double1}) turns out to be

\ba
{1 \over 2}(\langle S_I\rangle_f^2-\langle S_I^2\rangle_f) 
&=& -\frac{\lambda_1^2}{4} \int d\tilde R \int dr 
\Big[\Big\{-{\beta^2 \over 2\pi}dl\Big(1+ln\Big|{a \over r}
\Big|\Big)\Big\}\cos 2\beta[\Phi_{s\Lambda'}(\tilde R)]
\nonumber \\
& &+\Big\{{\beta^2 \over 2\pi}dl\Big(1+ln\Big|{a \over r}
\Big|\Big)\Big\}
\cos\beta [r\partial_{\tilde R}\Phi_{s\Lambda'}(\tilde R)]\Big] + \ldots
\label{order2a}
\ea
where we have expanded the cosines in $r$. 
The ellipsis represent the $O(\lambda_2^2)$ (which is identical to the 
$O(\lambda_1^2)$ part under replacement of $\beta$ by $2\beta$) and 
$O(\lambda_1\lambda_2)$ contributions. Following the arguments 
given in section 2, we reach to the following expression

\ba
{1 \over 2}(\langle S_I\rangle_f^2-\langle S_I^2\rangle_f) 
&=&{\lambda_1^2 \over 4} \Big[\int dr 
\Big\{{\beta^2 \over 2\pi}dl\Big(1+ln\Big|{a \over r}
\Big|\Big)\Big\}\Big]\int d\tilde R~\cos 2\beta[\Phi_{s\Lambda'}(\tilde R)]
\nonumber \\
& &-{\lambda_1^2 \over 4}\Big[\int dr
\Big\{{\beta^2 \over 2\pi}dl\Big(1+ln\Big|{a \over r} 
\Big|\Big)\Big\}\Big]\int d\tilde R~\bf 1 + \ldots\,,
\label{order2b}
\ea    
where the integration over $r$ gives rise to nonuniversal numerical
constant $I_1$ given by 

\be
I_1 = \frac{1}{8\pi}\int_{-a}^{a}dr~
\Bigg(1+ln\Big|\frac{a}{r}\Big|\Bigg)=\frac{a}{2\pi}\,. 
\label{I1}
\ee

Similarly use of center-of-mass and relative
coordinates $\tilde R$ and $r$ respectively and expansion of the cosines in
$r$ simplifies the $O(\lambda_1\lambda_2)$ contribution
to~(\ref{order2double1}) into

\ba
& &\frac{1}{2}(\langle S_I\rangle_f^2-\langle S_I^2\rangle_f) 
\nonumber \\
& &=\lambda_1\lambda_2\beta^2\Big(1-\frac{5\beta^2}{4\pi}dl\Big)\int
dr~G_f(r)\int d\tilde R~\Big(\cos\beta[3\Phi_{s\Lambda'}(\tilde R)
+\frac{r}{2}\partial_{\tilde R}\Phi_{s\Lambda'}(\tilde R)]
\nonumber \\
& &-\cos\beta[\Phi_{s\Lambda'}(\tilde 
R)+\frac{3r}{2}\partial_{\tilde R}\Phi_{s\Lambda'}(\tilde R)]\Big)+\ldots
\nonumber \\
& &=\lambda_1\lambda_2\beta^2\Big(1-\frac{5\beta^2}{4\pi}dl\Big)\int
dr~G_f(r)\int d\tilde R~\Big[\cos 3\beta\Phi_{s\Lambda'}(\tilde R)
\Big(1-\frac{\beta^2}{2}~\frac{r^2}{4}(\partial_{\tilde
R}\Phi_{s\Lambda'}(\tilde R))^2\Big)
\nonumber \\
& &-\sin 3\beta\Phi_{s\Lambda'}(\tilde
R)\cdot \frac{r}{2}\partial_{\tilde R}\Phi_{s\Lambda'}(\tilde R)
-\cos \beta\Phi_{s\Lambda'}(\tilde R)
\Big(1-\frac{\beta^2}{2}~\frac{9r^2}{4}(\partial_{\tilde
R}\Phi_{s\Lambda'}(\tilde R))^2\Big)
\nonumber \\
& &+\sin \beta\Phi_{s\Lambda'}(\tilde
R)\cdot \frac{3r}{2}\partial_{\tilde R}\Phi_{s\Lambda'}(\tilde R)\Big]
+\ldots\,,
\ea
where in the last step we expanded the cosines in $r$ about $\tilde
R$. By the same argument applied to $O(\lambda_1^2)$ part we can neglect the 
(gradient)$^2$ terms. As the field $\Phi$ is odd under $\tilde R$
reflection, the overall sine terms are not invariant under this
operation and hence do not contribute. Hence using~(\ref{fastgf1}) and
the nonuniversal constants resulting from the integrals

\ba
I_2&=&-\frac{1}{48\pi}\int_{-a}^{a}dr_1\int_{-a}^{a}dr_2~
ln\Bigg|\frac{(r_1-r_2)r_2}{r_1^2}\Bigg|=\frac{\pi a^2}{144}  \,, 
\nonumber \\
I_3&=&-\frac{1}{48\pi}\int_{-a}^{a}dr_1\int_{-a}^{a}dr_2~
ln\Bigg|\frac{(r_1-r_2)r_2^3}{r_1^4}\Bigg|=\frac{\pi a^2}{144}  \,,
\label{I2I3}
\ea
the $O(\lambda_1\lambda_2)$ contribution simplifies to 

\be
{1 \over 2}(\langle S_I\rangle_f^2-\langle S_I^2\rangle_f)
=2\frac{\tilde\lambda_1\tilde\lambda_2}{\pi}\beta^2dl
\int \frac{d\tilde R}{a}~(\cos
3\beta\Phi_{s\Lambda'}(\tilde R) 
-\cos\beta\Phi_{s\Lambda'}(\tilde R))+\ldots\,, 
\label{order2double-mixed}
\ee
neglecting the $O(dl^2)$ terms. Typically the integrals in~(\ref{I2I3})
are divergent. But we have regulated them excluding the origin which
is sensible in present case.

\vskip0.5cm

\underline{\bf Third order contribution:}

\vskip0.5cm

The cubic contribution to (\ref{SIeff}) reads

\ba
& &{1 \over 6}\langle S_I^3\rangle_f -{1 \over 2}\langle S_I\rangle_f
\langle S_I^2\rangle_f+{1 \over 3}\langle S_I\rangle_f^3
\nonumber \\
&=&-\frac{\lambda_1^3}{6}\int dt_1\int dt_2\int dt_3
\Bigg\{\Bigg(
\frac{\beta^2}{8\pi}dl~ln\Bigg|\frac{(t_2-t_3)(t_3-t_1)}{(t_1-t_2)^2}\Bigg|
\Bigg)
\cos\beta[\Phi_{s\Lambda'}(t_1)+\Phi_{s\Lambda'}(t_2)+\Phi_{s\Lambda'}(t_3)]
\nonumber \\
& &+\Bigg(
\frac{\beta^2}{8\pi}dl~\Bigg(
ln\Bigg|\frac{a^4}{(t_1-t_2)^2(t_2-t_3)(t_3-t_1)}\Bigg| 
+4\Bigg)
\Bigg)
\cos\beta[\Phi_{s\Lambda'}(t_1)+\Phi_{s\Lambda'}(t_2)-\Phi_{s\Lambda'}(t_3)]
\nonumber \\
& &+\Bigg(
\frac{\beta^2}{8\pi}dl~\Bigg(
ln\Bigg|\frac{(t_1-t_2)^2(t_3-t_1)}{a^2(t_2-t_3)}\Bigg| 
-2\Bigg)
\Bigg)
\cos\beta[\Phi_{s\Lambda'}(t_1)-\Phi_{s\Lambda'}(t_2)+\Phi_{s\Lambda'}(t_3)]
\nonumber \\
& &+\Bigg(
\frac{\beta^2}{8\pi}dl~\Bigg(
ln\Bigg|\frac{(t_1-t_2)^2(t_2-t_3)}{a^2(t_3-t_1)}\Bigg| 
-2\Bigg)
\Bigg)
\cos\beta[\Phi_{s\Lambda'}(t_1)-\Phi_{s\Lambda'}(t_2)
-\Phi_{s\Lambda'}(t_3)]\Bigg\} + \ldots\,,
\nonumber \\
& &
\label{order3double1}
\ea
neglecting the $O(dl^2)$ terms. The ellipsis represent 
the $O(\lambda_2^3)$, $O(\lambda_1^2\lambda_2)$, and  
$O(\lambda_1\lambda_2^2)$ contributions. The $O(\lambda_2^3)$
and $O(\lambda_1\lambda_2^2)$ are related to $O(\lambda_1^3)$
and $O(\lambda_1^2\lambda_2)$ by exchanging $\beta \leftrightarrow 2\beta$
respectively.
We again introduce center-of-mass coordinate $\bar R =
(t_1+t_2+t_3)/3$ and relative coordinates $r_1=t_2-t_1$ and
$r_2=t_3-t_1$. The relative coordinates being bounded within the UV
cut-off $r_1,r_2<\Lambda^{-1} \sim a$, the Green's functions are
truly short-ranged and we can expand the cosines in $r_1,r_2$.
The result is 

\ba
& &{1 \over 6}\langle S_I^3\rangle_f
-{1 \over 2}\langle S_I\rangle_f\langle S_I^2\rangle_f
+{1 \over 3}\langle S_I\rangle_f^3
\nonumber \\
&=&-\lambda_1^3\frac{\beta^2}{48\pi}dl\Bigg[\int dr_1\int
dr_2~ln\Big|\frac{(r_1-r_2)r_2}{r_1^2}\Big|\int d\bar
R~\cos 3\beta\Phi_{s\Lambda'}(\bar R) 
\nonumber \\
& &+ \int dr_1\int dr_2~(4+4~ln|a|-ln|r_1^2r_2(r_1-r_2)|) 
\nonumber \\
& &\cdot\int d\bar R~\cos\beta[\Phi_{s\Lambda'}(\bar 
R)-\frac{4}{3}r_1\partial_{\bar R}\Phi(\bar
R,r_2)+\frac{2}{3}r_2\partial_{\bar R}\Phi_{s\Lambda'}(\bar R,r_1)]
\nonumber \\
& &+ \int dr_1\int
dr_2~(-4-4~ln|a|+2~ln\Big|\frac{r_1^2(r_1-r_2)}{r_2}\Big|) 
\nonumber \\
& &\cdot\int d\bar R~\cos\beta[\Phi_{s\Lambda'}(\bar R)
-\frac{4}{3}r_2 \partial_{\bar R}\Phi_{s\Lambda'}(\bar 
R,r_1)+\frac{2}{3}r_1\partial_{\bar R}\Phi_{s\Lambda'}(\bar R,r_2)]\Bigg]
+\ldots\,.
\label{order3a}
\ea
Notice that the expression 

\ba
& &\int dr_1\int dr_2~(4+4~ln|a|)\cos\beta[\Phi_{s\Lambda'}(\bar 
R)-\frac{4}{3}r_1\partial_{\bar R}\Phi_{s\Lambda'}(\bar
R,r_2)+\frac{2}{3}r_2\partial_{\bar R}\Phi_{s\Lambda'}(\bar R,r_1)]
\nonumber \\
& &+ \int dr_1\int
dr_2~(-4-4~ln|a|)\cos\beta[\Phi_{s\Lambda'}(\bar R)- \frac{4}{3}r_2
\partial_{\bar R}\Phi_{s\Lambda'}(\bar 
R,r_1)+\frac{2}{3}r_1\partial_{\bar R}\Phi_{s\Lambda'}(\bar R,r_2)]
\nonumber \\
& &
\ea 
vanishes identically under the exchange of the dummy variables $r_1,r_2$.
Hence the cubic contribution simplifies to

\ba
& &{1 \over 6}\langle S_I^3\rangle_f
-{1 \over 2}\langle S_I\rangle_f\langle S_I^2\rangle_f+{1 \over 3}
\langle S_I\rangle_f^3
\nonumber \\
&=&-\lambda_1^3\frac{\beta^2}{48\pi}dl
\Bigg[\int dr_1\int
dr_2~ln\Big|\frac{(r_1-r_2)r_2}{r_1^2}\Big|\int d\bar
R~\cos3\beta\Phi_{s\Lambda'}(\bar R) 
+ \int dr_1\int dr_2~ln\Big|\frac{r_2^3(r_1-r_2)}{r_1^4}\Big|
\nonumber \\
& &\cdot\int d\bar R~\cos\beta[\Phi_{s\Lambda'}(\bar 
R)-\frac{4}{3}r_1\partial_{\bar R}\Phi_{s\Lambda'}(\bar
R,r_2)+\frac{2}{3}r_2\partial_{\bar R}\Phi_{s\Lambda'}(\bar R,r_1)]\Bigg]
+\ldots\,.
\label{order3b}
\ea
Since the relative coordinates $r_1,r_2$ are small, the second cosine
simplifies to

\ba
& &\cos\beta[\Phi_s(\bar 
R)-\frac{4}{3}r_1\partial_{\bar R}\Phi_s(\bar
R,r_2)+\frac{2}{3}r_2\partial_{\bar R}\Phi_s(\bar R,r_1)]
\nonumber \\
&\approx & \cos\beta\Phi_s(\bar
R)\Bigg(1-\frac{\beta^2}{2}\bigg(\frac{2}{3}r_2\partial_{\bar R}\Phi_s(\bar
R,r_1)-\frac{4}{3}r_1\partial_{\bar R}\Phi_s(\bar R,r_2)\bigg)^2\Bigg)
\nonumber \\
& &-\bigg(\frac{2}{3}r_2\partial_{\bar R}\Phi_s(\bar
R,r_1)-\frac{4}{3}r_1\partial_{\bar R}\Phi_s(\bar
R,r_2)\bigg)\sin\beta\Phi_s(\bar R)\,.
\nonumber \\
& &
\ea 
Substituting this into (\ref{order3a}) we get

\ba
& &{1 \over 6}\langle S_I^3\rangle_f-{1 \over 2}\langle S_I\rangle_f
\langle S_I^2\rangle_f+{1 \over 3}\langle S_I\rangle_f^3
\nonumber \\
&=&-\lambda_1^3\frac{\beta^2}{48\pi}dl
\Bigg[\int dr_1\int
dr_2~ln\Big|\frac{(r_1-r_2)r_2}{r_1^2}\Big|\int d\bar
R~\cos3\beta\Phi_{s\Lambda'}(\bar R) 
\nonumber \\
& &+ \int dr_1\int dr_2~ln\Big|\frac{r_2^3(r_1-r_2)}{r_1^4}\Big|
\int d\bar R~\cos\beta\Phi_{s\Lambda'}(\bar R)
\nonumber \\
& &-\frac{\beta^2}{2}\int dr_1\int
dr_2~ln\Big|\frac{r_2^3(r_1-r_2)}{r_1^4}\Big| 
\Big(\frac{4}{9}r_2^2+\frac{16}{9}r_1^2-\frac{16}{9}r_1r_2\Big)\int
d\bar R~\cos\beta\Phi_{s\Lambda'}(\bar R) \bigg(\partial_{\bar
R}\Phi_{s\Lambda'} (\bar R)\bigg)^2 
\nonumber \\
& &- \int dr_1\int dr_2\int d\bar
R~ln\Big|\frac{r_2^3(r_1-r_2)}{r_1^4}\Big|
\bigg(\frac{2}{3}r_2\partial_{\bar R}\Phi_{s\Lambda'}(\bar
R,r_1)-\frac{4}{3}r_1\partial_{\bar R}\Phi_{s\Lambda'}(\bar R,r_2)\bigg)
\sin\beta\Phi_{s\Lambda'}(\bar R)\Bigg]+\ldots\,.
\nonumber \\
& &
\label{order3c}
\ea
Since under reflection of $\bar R$ the integrand containing sine is
odd, it vanishes identically. The $\bar R$ integration  containing
$\cos\beta\Phi(\bar R)\bigg(\partial_{\bar R}\Phi (\bar R)\bigg)^2$ 
vanishes imposing equation of motion from the action where the bulk
degrees of freedom have been integrated out.
Hence the expression for $O(\lambda_1^3)$ cubic contribution is

\ba
& &{1 \over 6}\langle S_I^3\rangle_f-{1 \over 2}\langle S_I\rangle_f
\langle S_I^2 \rangle_f+{1 \over 3}\langle S_I\rangle_f^3
\nonumber \\
&=&-\lambda_1^3\frac{\beta^2}{48\pi}dl
\Bigg[\int dr_1\int
dr_2~ln\Big|\frac{(r_1-r_2)r_2}{r_1^2}\Big|\int d\bar
R~\cos3\beta\Phi_{s\Lambda'}(\bar R) 
\nonumber \\
& &+ \int dr_1\int dr_2~ln\Big|\frac{r_2^3(r_1-r_2)}{r_1^4}\Big|
\int d\bar R~\cos\beta\Phi_{s\Lambda'}(\bar R)\Bigg]+\ldots\,,
\label{order3d}
\ea
where the integrations over $r_1,r_2$ give rise to nonuniversal 
constants given by~(\ref{I2I3}).

Next we turn to $O(\lambda_1^2\lambda_2)$ part. Using~(\ref{exp-cor}), 
(\ref{fastgf}) and~(\ref{fastgf1}) and neglecting $O(dl^2)$ terms it
simplifies to

\ba
& &\frac{1}{6}\langle S_I^3\rangle_f-{1 \over 2}\langle S_I\rangle_f
\langle S_I^2\rangle_f+{1 \over 3}\langle S_I\rangle_f^3
\nonumber \\
&=&\frac{1}{8}\lambda_1^2\lambda_2\Big(1-\frac{3\beta^2}{2\pi}dl\Big)
\int dt_1\int dt_2\int dt_3
\Bigg[
\frac{\beta^2}{\pi}dl~ln\Big|\frac{t_3-t_1}{t_3-t_2}\Big|\cos \beta
(\Phi_{s\Lambda'}(t_1)+\Phi_{s\Lambda'}(t_2)+2\Phi_{s\Lambda'}(t_3))
\nonumber \\
& &-\frac{\beta^2}{\pi}dl~ln\Big|\frac{t_3-t_1}{t_3-t_2}\Big|\cos \beta
(-\Phi_{s\Lambda'}(t_1)-\Phi_{s\Lambda'}(t_2)+2\Phi_{s\Lambda'}(t_3))
\nonumber \\
& &-\Bigg(\frac{\beta^2}{\pi}dl~ln\Big|\frac{a^2}{(t_3-t_2)(t_3-t_1)}\Big|
+\frac{2\beta^2}{\pi}dl\Bigg)\cos\beta
(\Phi_{s\Lambda'}(t_1)-\Phi_{s\Lambda'}(t_2)+2\Phi_{s\Lambda'}(t_3))
\nonumber \\
& &+\Bigg(\frac{\beta^2}{\pi}dl~ln\Big|\frac{a^2}{(t_3-t_2)(t_3-t_1)}\Big|
+\frac{2\beta^2}{\pi}dl\Bigg)\cos\beta
(-\Phi_{s\Lambda'}(t_1)+\Phi_{s\Lambda'}(t_2)+2\Phi_{s\Lambda'}(t_3))
\Bigg]+\ldots
\nonumber \\
& &
\ea
As in the previous case we introduce the center-of-mass coordinate
$\bar R=(t_1+t_2+t_3)/3$ and relative coordinates $r_1=t_3-t_1$ and
$r_2=t_3-t_2$. Expanding the cosines in $r_1,r_2$, the above
expression can be written approximately as 

\be
\frac{1}{8}\frac{\beta^2}{\pi}dl\int dr_1\int dr_2
~(ln~r_1-ln~r_2) \cdot \int d\bar R (\cos 4\beta\Phi_{s\Lambda'}(\bar
R)-\bf 1)
\ee
which vanishes identically after integration over $r_1$ and
$r_2$. By doing similar analysis for $O(\lambda_1\lambda_2^2)$, we
conclude that all
the contributions to $O(\lambda_1^2\lambda_2)$ and
$O(\lambda_1\lambda_2^2)$ vanishes.



\end{document}